\documentclass[structabstract]{raa}
\usepackage{graphicx,times}
\usepackage{natbib,amssymb}
\usepackage{epsfig}
\usepackage{float}

\providecommand{\bm}[1]{\mbox{\boldmath$#1$\unboldmath}}

\begin{document}
   \title{Mass-accreting white dwarfs and type Ia supernovae}

   \volnopage{ {\bf 2018} Vol.\ {\bf XX} No. {\bf XXX}, 000--000}
   \setcounter{page}{1}

   \author{Bo Wang \inst{1,2,3}
          }
   \institute{Key Laboratory for the Structure and Evolution of Celestial Objects, Yunnan Observatories, Chinese Academy of Sciences,  Kunming 650216, China;
          {\it wangbo@ynao.ac.cn}\\
          \and
          Center for Astronomical Mega-Science, Chinese Academy of Sciences, Beijing 100012, China
          \and
          University of Chinese Academy of Sciences, Beijing 100049, China
              }
              

   \date{Received; accepted}

\abstract {Type Ia supernovae (SNe~Ia)  play a prominent role in understanding the evolution of Universe. 
They are thought  to be thermonuclear explosions of 
mass-accreting carbon-oxygen white dwarfs (CO WDs) in binaries,
although the mass donors of the accreting WDs are still not well determined. 
In this article, I review recent studies on  mass-accreting WDs, including H- and He-accreting WDs. 
I also review currently most studied  progenitor models of SNe Ia, i.e.,
the single-degenerate model (including  the WD+MS channel, the WD+RG channel  and  the WD+He star channel), 
the double-degenerate model (including the violent merger scenario) 
and the sub-Chandrasekhar mass model. Recent progress on these progenitor models is discussed, 
including the initial parameter space for producing SNe Ia, 
the binary evolutionary paths to SNe Ia,  the progenitor candidates of SNe Ia, 
the possible surviving companion stars of SNe Ia,
and some observational constraints, etc. 
Some other potential progenitor models of SNe Ia are also summarized, including the hybrid CONe WD model,
the core-degenerate model, the double WD  collision model, the spin-up/spin-down model, and the model of WDs near black holes.
To date, it seems that two or more progenitor models are needed  to explain the  observed diversity among SNe Ia.
 }
\titlerunning{Mass-accreting WDs and SNe Ia}
\authorrunning{B. Wang }
\maketitle


\section{Introduction}

Type Ia supernovae (SNe Ia)  are defined as SNe with strong SiII absorption lines in their spectra,
but without H and He lines  nearby their maximum  luminosity (see Filippenko 1997).
They happen in all kinds of galaxies, including young and old stellar populations  (e.g. Branch et al. 1993; Wang et al. 1997).
SNe Ia are accurate distance measurements in cosmology due to 
the uniformity of their light curves, revealing the accelerating expansion of
the current Universe  driven by dark energy (e.g., Riess et al. 1998; Perlmutter et al.
1999; Howell 2011).  They are  element factories in the chemical evolution of galaxies, which
are the main producer of iron to their host galaxies (e.g., Greggio \& Renzini 1983; Matteucci \& Greggio 1986).
They are also the sources of kinetic energy in galaxy evolution, the accelerators of cosmic rays,  
and the endings of binary evolution (e.g., Helder et al. 2009; Powell et al. 2011; Fang \& Zhang 2012).  

The Phillips relation is adopted when SNe Ia are applied as distance indicators, 
which is a width-luminosity relation among SNe Ia; events with wider light curves are brighter (see Phillips 1993; Phillips et al. 1999).
However,  more and more observational evidence indicates that  there exists spectroscopic diversity among SNe Ia and
not all SNe Ia obey the Phillips relation (e.g., Li et al. 2001, 2011a; Wang et al. 2006; Branch et al. 2009;   
Foley et al. 2009, 2018; Blondin et al. 2012;
Zhang et al. 2014, 2016; Zhai et al. 2016; Taubenberger 2017).
The light curves of SNe Ia are powered by
the radioactive decay of $^{56}$Ni $\rightarrow$ $^{56}$Co
$\rightarrow$ $^{56}$Fe (e.g., Arnett 1982; Branch \& Tammann 1992).

It has been suggested that some stellar parameters
at the moment of SN explosion may affect the final amount of $^{56}$Ni,
and thus the maximum light of SNe Ia, for example,  
the metallicity (e.g., Timmes et al. 2003; Podsiadlowski et al. 2006;  Sullivan et al. 2010; Bravo et al. 2010),
the average ratio of carbon to oxygen  of a WD  (e.g., Umeda et al. 1999), 
and the transition density  from deflagration to detonation or the number of ignition points in the center of WDs 
(e.g., Hillebrandt \& Niemeyer 2000;  Kasen et al. 2009; H\"{o}flich et al. 2010), etc.
Maeda et al. (2010) argued that the observed
SN Ia diversity may be a result of off-center ignition coupled 
with the observer's viewing angle (see also Parrent et al. 2011).
Note that Meng et al. (2017) recently argued that all kinds of SNe Ia may obey the same polarization sequence that might be 
explained by the delayed-detonation explosion model.

SNe Ia are thought to be outcomes of thermonuclear  explosions of 
mass-accreting carbon-oxygen white dwarfs (CO WDs)  that have mass close to
the Chandrasekhar limit (${M}_{\rm Ch}$; e.g., Hoyle \& Fowler 1960; Nomoto et al. 1984).
The  WD explosion  with ${M}_{\rm Ch}$ can  
reproduce  the observed light curves and spectroscopy of  most SNe Ia
(e.g., H\"{o}flich et al.1996;  Podsiadlowski et al. 2008; Leung \& Nomoto 2017),
and most SNe Ia are inferred to have total ejecta masses close to ${M}_{\rm Ch}$ (see Mazzali et al. 2007). 
Umeda et al. (1999)  suggested that the birth mass of a CO WD is usually  $<$1.1\,${M}_\odot$ (see also Siess 2006; Doherty et al. 2015, 2017),
and thus a CO WD needs to obtain enough mass from its companion in a binary before it explodes as an SN Ia.
However, the nature of the companion of the CO WD is still not well determined over the 
past 60 years of SN research though there exist many observational constraints
(e.g.,  Mannucci et al. 2006; F\"{o}rster et al. 2006, 2013; Aubourg et al. 2008; Maoz et al. 2011; Wang et al. 2013a;
Graur \& Maoz 2013; Mart\'{i}nez-Rodr\'{i}guez et al. 2017; Heringer et al. 2017), 
which involves the progenitor issue  of SNe Ia (for recent reviews see Wang \& Han 2012; Maoz \& Mannucci 2012; H\"{o}flich et al. 2013;  
Hillebrandt et al. 2013;  Maoz et al. 2014; Ruiz-Lapuente 2014; Parrent et al. 2014; Maeda \& Terada 2016; Branch \& Wheeler 2017; Soker  2018).

Many progenitor models  have been proposed to explain the  observed diversity among SNe Ia, in which
the most studied models are the single-degenerate (SD) model,
the double-degenerate (DD) model  and the  sub-${M}_{\rm Ch}$ model.
(1) \textit{The SD model}.
In this model, a CO WD accretes H-/He-rich  material from a non-degenerate donor. The WD may produce an SN Ia when it 
grows in mass close to ${M}_{\rm Ch}$ (e.g., Whelan \& Iben 1973; Nomoto et al.1984).
(2) \textit{The DD model}. 
In this model, a CO WD merges with another CO WD, the merging of which is due to the gravitational wave radiation, 
producing an SN Ia finally (e.g., Webbink 1984; Iben \& Tutukov 1984).
(3) \textit{The sub-${M}_{\rm Ch}$ model}. 
In this model, the thermonuclear explosion of a CO WD  results from the detonation at the bottom of  a He-shell, in which
the CO WD has the mass below ${M}_{\rm Ch}$ (e.g., Nomoto 1982a; Woosley et al. 1986). 

In this article,  I mainly review recent studies on mass-accreting WDs and different progenitor models of SNe Ia. 
In Sect. 2,  I review recent studies of  H- and He-accreting WDs in detail. 
I  also review recent progress on  the currently most discussed progenitor models  of SNe Ia,  including
the SD model in Sect. 3, the DD model in Sect. 4, and the sub-${M}_{\rm Ch}$ model in Sect. 5.
In Sect. 6, I summarize some other potential progenitor models of SNe Ia.
Finally, a summary is given in Sect. 7. 
For more discussions on the progenitors, explosion mechanisms 
and observational properties of SNe Ia,
see previous reviews, e.g., Branch et al. (1995), Nomoto et al. (1997), 
Hillebrandt \& Niemeyer (2000), Livio (2000), Wang \& Wheeler (2008) and Podsiadlowski (2010).

\section{Mass-accreting white dwarfs}

A WD  in a binary system can usually accrete H-/He-rich material from its mass donor. 
The process of mass-accretion onto WDs 
is important for the studies of binary evolution and accretion physics.
Employing the  stellar evolution code called
Modules for Experiments in Stellar Astrophysics (MESA; see Paxton et al. 2011, 2013, 2015), 
Wang et al. (2015a) recently studied the long-term evolution of the He-accreting WDs with 
various initial WD masses (${M}^{\rm i}_{\rm WD}=0.5-1.35\,{M}_\odot$) and 
accretion rates ($\dot{M}_{\rm acc}=10^{-8}-10^{-5}\,M_\odot\,\mbox{yr}^{-1}$).
The super-Eddington wind is supposed as the mass-loss mechanism during He-shell flashes (e.g., Denissenkov et al. 2013; Ma et al. 2013). 
The initial WD models in Wang et al. (2015a) have a metallicity of 2\%, 
and the accreted He-rich material consists of 98\% He and 2\% metallicity.
In this article, I simulated the long-term evolution of the H-accreting WDs with 
various ${M}^{\rm i}_{\rm WD}$ and $\dot{M}_{\rm acc}$ using MESA (version 7624), 
in which the accreted H-rich material consists of 70\% H, 28\% He and 2\% metallicity.
Basic assumptions and input here are similar to those of Wang et al. (2015a).
In my computations,  the WDs were resolved with $>$2000 meshpoints.

\subsection{Stable burning regime}

\begin{figure*}
\begin{center}
\includegraphics[width=10.2cm,angle=0]{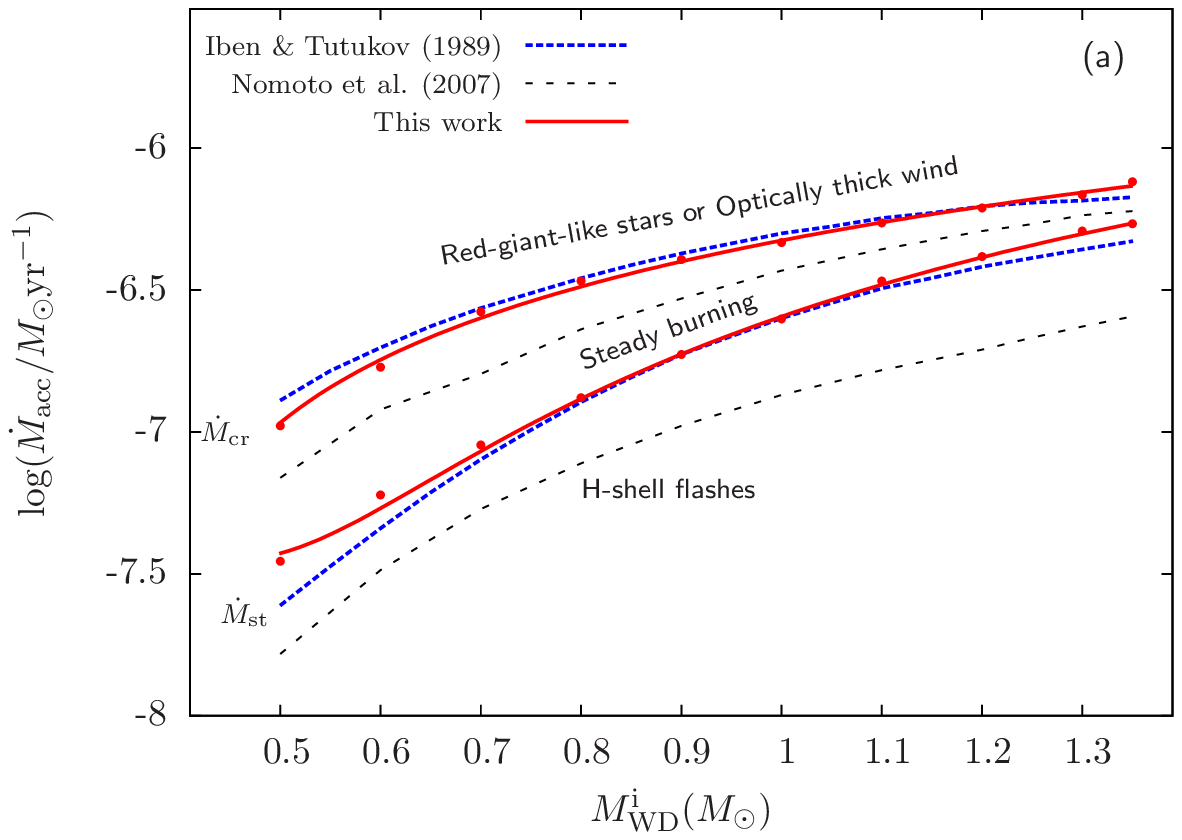}
\includegraphics[width=10.2cm,angle=0]{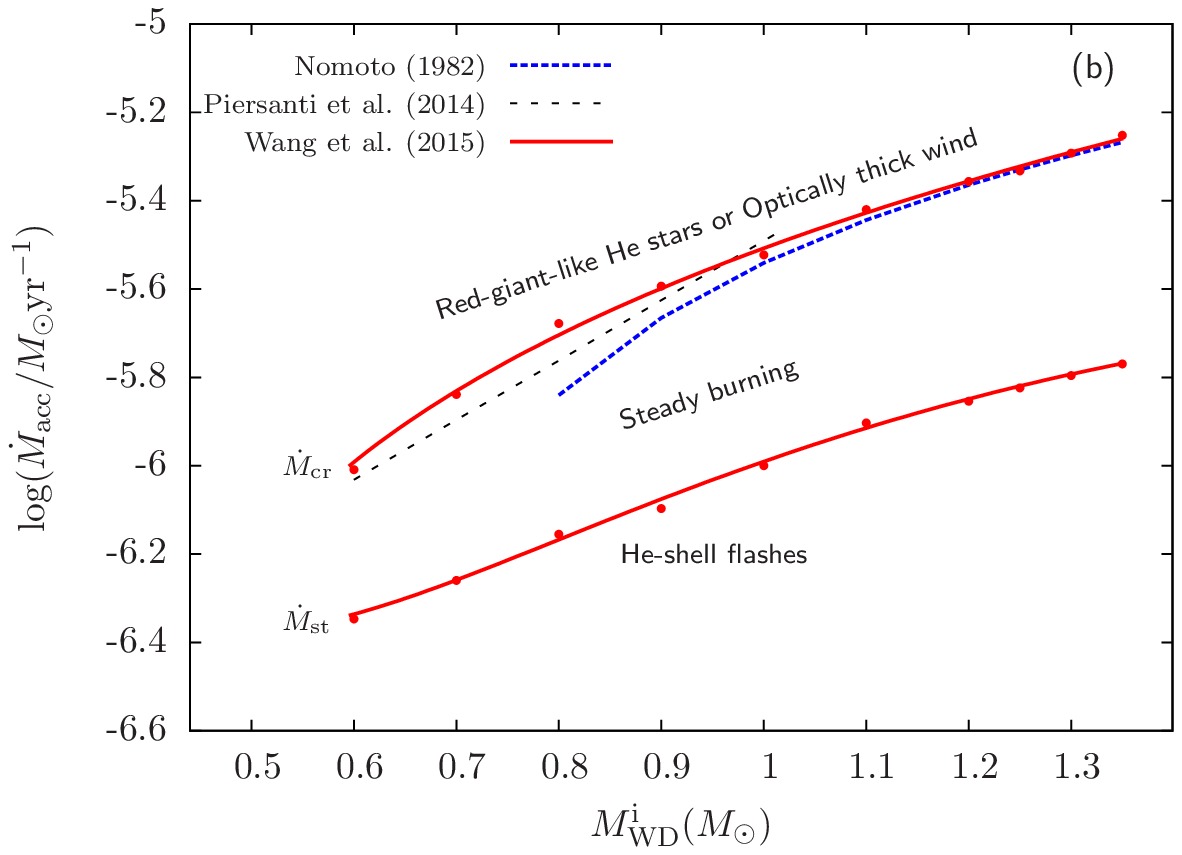}
\caption{Stable H-/He-shell burning regime in the ${M}^{\rm i}_{\rm WD}-\dot{M}_{\rm acc}$
plane. In panel (a), it shows the stable H-shell burning regime. 
The red solid lines are the results of my simulations,  the blue dotted lines are taken from Iben \& Tutukov (1989), and the black dashed lines are from Nomoto et al. (2007).
In panel (b), it presents the stable He-shell burning regime. 
The red solid lines are taken from Wang et al. (2015a), 
the blue dotted line is taken from Nomoto (1982b), and the black dashed line is from Piersanti et al. (2014). }
\end{center}
\end{figure*}

In Fig. 1, I show the stable H-/He-shell burning regime in the ${M}^{\rm i}_{\rm WD}-\dot{M}_{\rm acc}$
plane.   In this steady burning regime,
WD binaries have been identified as  supersoft X-ray sources in the observations (e.g.,  van den Heuvel et al. 1992).

In the case of H-accreting WDs (see Fig. 1a), 
the final fate of  mass-accreting WDs is mainly determined by ${M}^{\rm i}_{\rm WD}$ and $\dot{M}_{\rm acc}$. 
If  $\dot{M}_{\rm acc}$ is larger than  the maximum accretion rate $\dot{M}_{\rm cr}$ for stable H-shell burning,
the WD will expand to red-giant dimensions and form a red-giant-like star due to the continuous pileup of the accreted material on its surface or, alternatively, 
the red-giant-like regime  is replaced by the optically thick wind regime (for more discussions see Sect. 3).
If $\dot{M}_{\rm acc}$ is below the minimum accretion rate $\dot{M}_{\rm st}$ for stable H-shel burning,  
the WD will experience multicycle H-shell flashes like
nova outbursts due to unstable nuclear burning.
The values of  $\dot{M}_{\rm cr}$ and   $\dot{M}_{\rm st}$  for  H-accreting WDs
can be approximated by the following formula
\begin{equation}
\small{\dot{M}_{\rm cr}=0.27\times10^{-7}({M}_{\rm WD}^2+25.52{M}_{\rm WD}-9.02)},
\end{equation}
\begin{equation}
\small{\dot{M}_{\rm st}=2.93\times10^{-7}(-{M}_{\rm WD}^3+4.41{M}_{\rm WD}^2-3.38{M}_{\rm WD}+0.84)},
\end{equation}
where ${M}_{\rm WD}$ is in units of ${M}_\odot$, and  $\dot{M}_{\rm cr}$ and $\dot{M}_{\rm st}$ are in units of $M_\odot\,\mbox{yr}^{-1}$.
I also compared my results with previous investigations of Iben \& Tutukov (1989) and Nomoto et al. (2007). 
It seems that my results are almost coincident with those of Iben \& Tutukov (1989),
but have some differences with those of Nomoto et al. (2007),  probably resulted from different methods adopted.
Nomoto et al. (2007) studied the mass-accretion process through a linear stability analysis, whereas I carried out
a detailed stellar evolution computations. 

In the case of  He-accreting WDs (see Fig. 1b), the final fate of the WDs is also determined by $\dot{M}_{\rm acc}$ and ${M}_{\rm WD}$.
The values of $\dot{M}_{\rm cr}$ and  $\dot{M}_{\rm st}$  are given as below
\begin{equation}
\small{\dot{M}_{\rm cr}=2.17\times10^{-6}({M}_{\rm WD}^2+0.82{M}_{\rm WD}-0.38)},
\end{equation}
\begin{equation}
\small{\dot{M}_{\rm st}=1.46\times10^{-6}(-{M}_{\rm WD}^3+3.45{M}_{\rm WD}^2-2.60{M}_{\rm WD}+0.85)}.
\end{equation}
It has been assumed that a WD can grow in mass to ${M}_{\rm Ch}$ in this stable He-shell burning regime 
and then explodes as an SN Ia (e.g., Nomoto 1982b; Wang et al. 2009a).
However,  Wang et al. (2017a) recently found that off-center carbon ignition happens on the surface of the WD 
if $\dot{M}_{\rm acc}$ is larger than a critical value  ($\sim$$2.05\times 10^{-6}\,{M}_\odot\,\mbox{yr}^{-1}$).
An off-center carbon ignition will convert CO WDs to ONe WDs via an inwardly propagating carbon burning flame;
ONe WDs are expected to collapse into
 a neutron star through electron capture on $^{24}$Mg and $^{20}$Ne when mass accretion goes on 
(e.g., Nomoto \& Iben 1985; Saio \& Nomoto 1985, 1998; Brooks et al. 2016; Wu \& Wang 2018).   
Wang et al. (2017a) found that the WD can increase its mass steadily in the regime between
$\dot{M}_{\rm st}$ and   the critical rate for off-center carbon burning, in which explosive carbon ignition 
(see Lesaffre et al. 2006; Chen et al. 2014c) can happen in the center
of the WD when it grows in mass close to ${M}_{\rm Ch}$, leading to an SN Ia explosion.  
Note that  Brooks et al. (2016) recently also reported these 
two possible outcomes (i.e., center or off-center carbon ignition), 
but they only computed over a narrower range of binary parameter space. 

Similar to previous studies (e.g., Iben \& Tutukov 1989; Nomoto et al. 2007), 
the He-shell burning underneath the H-shell was neglected for simplicity 
when I simulated the long-term evolution of  H-accreting WDs.
It is still hard for the H-accreting WD to increase its mass to ${M}_{\rm Ch}$ 
as steady burning regime of He-shell burning is higher than that for H-shell burning (see Fig. 1). 
This fundamental difficulty for double-shell (H-/He-shell) burning needs to be settled in future investigations.

\subsection{Mass-accumulative efficiencies and nova cycle durations}

If $\dot{M}_{\rm acc}<\dot{M}_{\rm st}$,   the accreting WD will experience H-/He-shell flashes like nova outbursts. 
Recent studies indicate that  a WD can grow in mass to  ${M}_{\rm Ch}$ through multicycle nova outbursts, 
resulting in an SN Ia explosion finally (e.g., Wang et al. 2015a; Hillman et al. 2015, 2016; Wu et al. 2017).

\begin{figure}
\begin{center}
\includegraphics[width=10.2cm,angle=0]{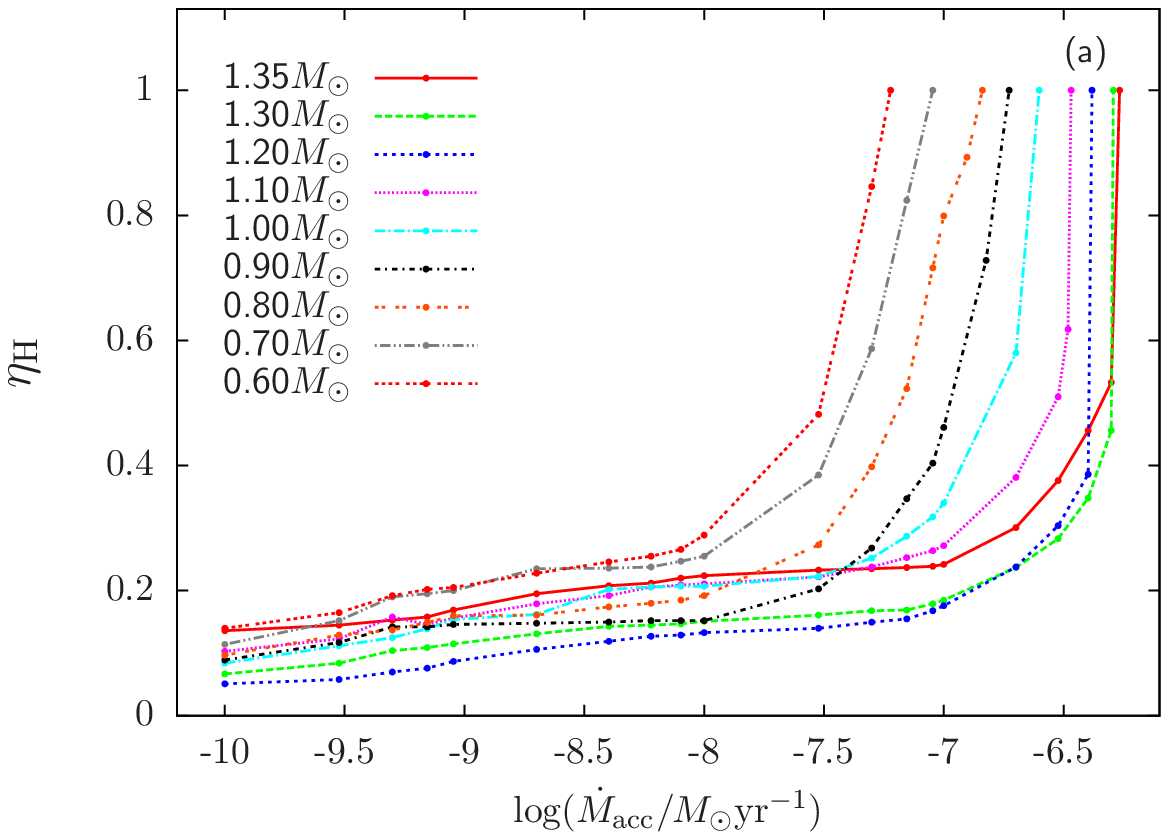}
\includegraphics[width=10.2cm,angle=0]{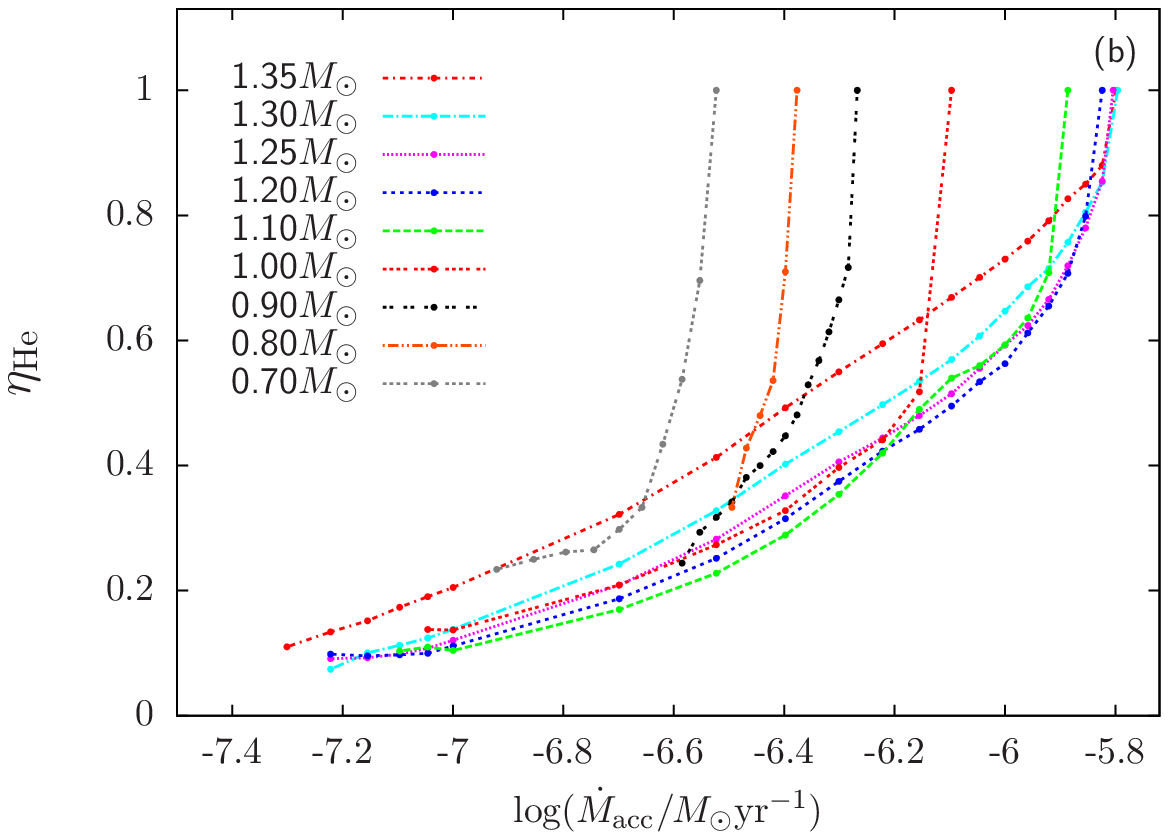}
\caption{Mass-accumulative efficiencies ($\eta$) vs. $\dot{M}_{\rm acc}$ for various ${M}^{\rm i}_{\rm WD}$. 
Panel (a): the results of H-shell flashes in my simulations.  
Panel (b): the results of He-shell flashes based on the studies of Wu et al. (2017).}
\end{center}
\end{figure}

The mass-accumulative efficiency ($\eta$) during nova outbursts  is  defined as
the mass fraction of accreted material that is retained by the WD.
$\eta$ plays a fundamental role in binary evolution, which has a strong influence 
on the rates and delay times of SNe Ia (see, e.g., Bours et al. 2013; Toonen et al. 2014; Wang et al. 2015b; Kato et al. 2018). 
Fig. 2 shows the mass-accumulative efficiencies  of H-/He-shell flashes for different ${M}^{\rm i}_{\rm WD}$ and $\dot{M}_{\rm acc}$. 
For a given ${M}_{\rm WD}$, $\eta$ increases with $\dot{M}_{\rm acc}$.This is because the degeneracy of 
the H-/He-shell is lower for high accretion rates, 
resulting in that the wind becomes weaker and more mass accumulated on the surface.
Yoon et al. (2004) suggested that $\eta$ may be increased when  rotation is considered. 
The data points of Fig. 2 can be used in the studies of binary population synthesis (BPS) computations, 
which can be provided on request by contacting the author.

\begin{figure}
\begin{center}
\includegraphics[width=9.5cm,angle=0]{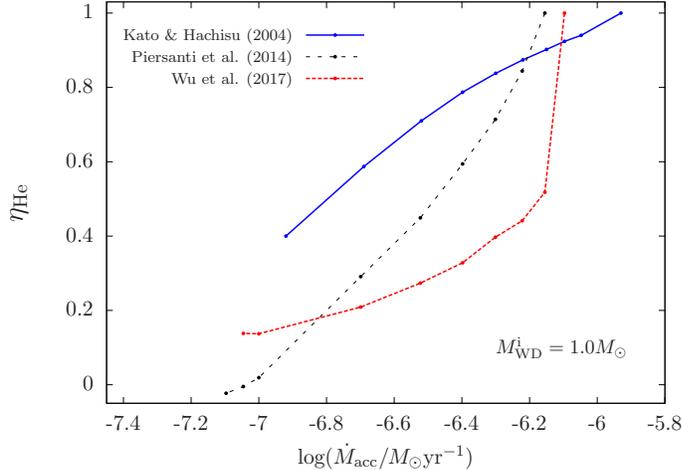}
 \caption{Mass-accumulative efficiencies ($\eta_{\rm He}$) vs. $\dot{M}_{\rm acc}$ 
 for ${M}^{\rm i}_{\rm WD}=1.0{M}_\odot$.
The blue solid line is  taken from Kato \& Hachisu (2004), 
the black dashed line is from Piersanti et al. (2014), and
the red dotted line is from Wu et al. (2017).}
  \end{center}
\end{figure}

Many studies on the long-term evolution of mass-accreting WDs  come 
into some different results about the value of $\eta$  (e.g., 
Prialnik \& Kovetz 1995;  Cassisi et al. 1998; Kato \& Hachisu 2004; Yaron et al. 2005;
Wolf et al. 2013; Idan et al. 2013; Newsham et al. 2014; Wang et al. 2015a; 
Hillman et al. 2015, 2016; Wu et al. 2017; Kato et al. 2017). 
In Fig. 3,  I compare the values of $\eta_{\rm He}$ obtained by 
different groups. From this figure, we can see that
the values of $\eta_{\rm He}$ in Kato \& Hachisu (2004) are apparently
higher than those in Piersanti et al. (2014) and Wu et al. (2017).
Kato et al. (2018) recently discussed the reasons for such divergence in detail, and 
found that the mass-loss mechanism during nova outbursts is a key process of 
determining the value of the mass-accumulative efficiency.

\begin{figure}
\begin{center}
\includegraphics[width=10.2cm,angle=0]{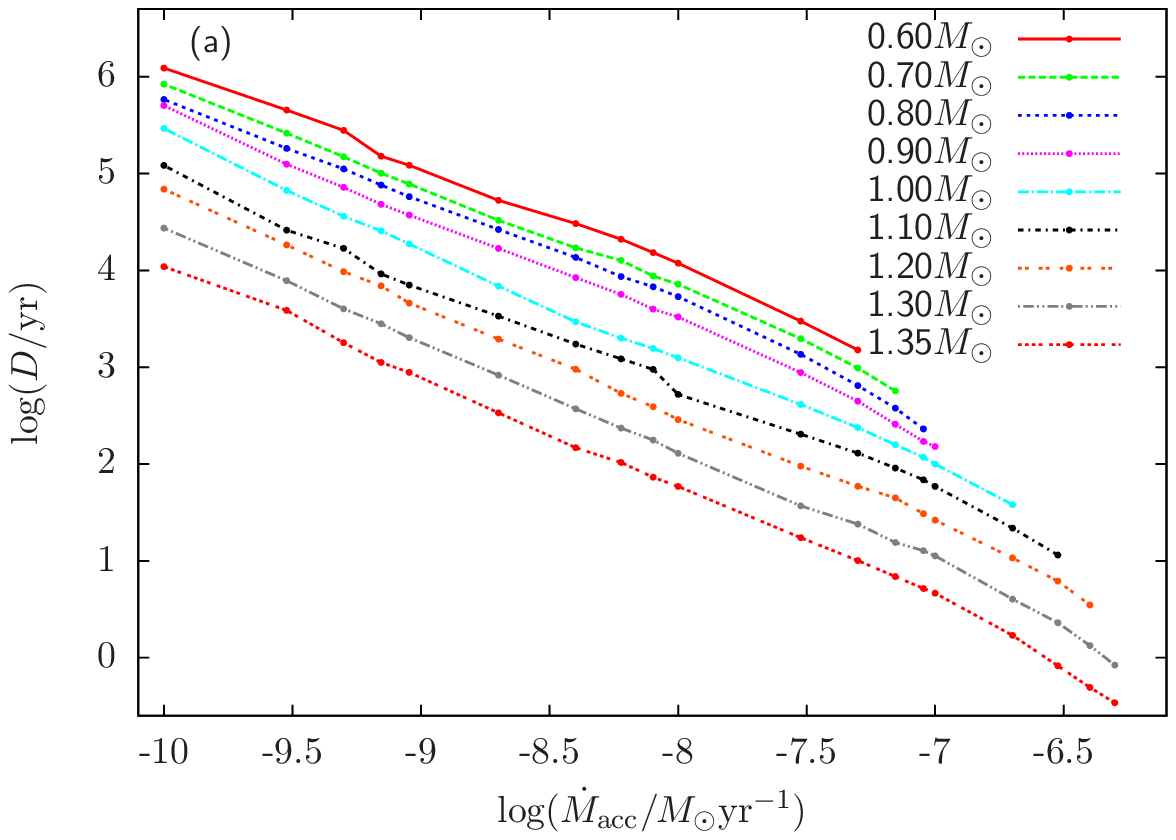}
\includegraphics[width=10.2cm,angle=0]{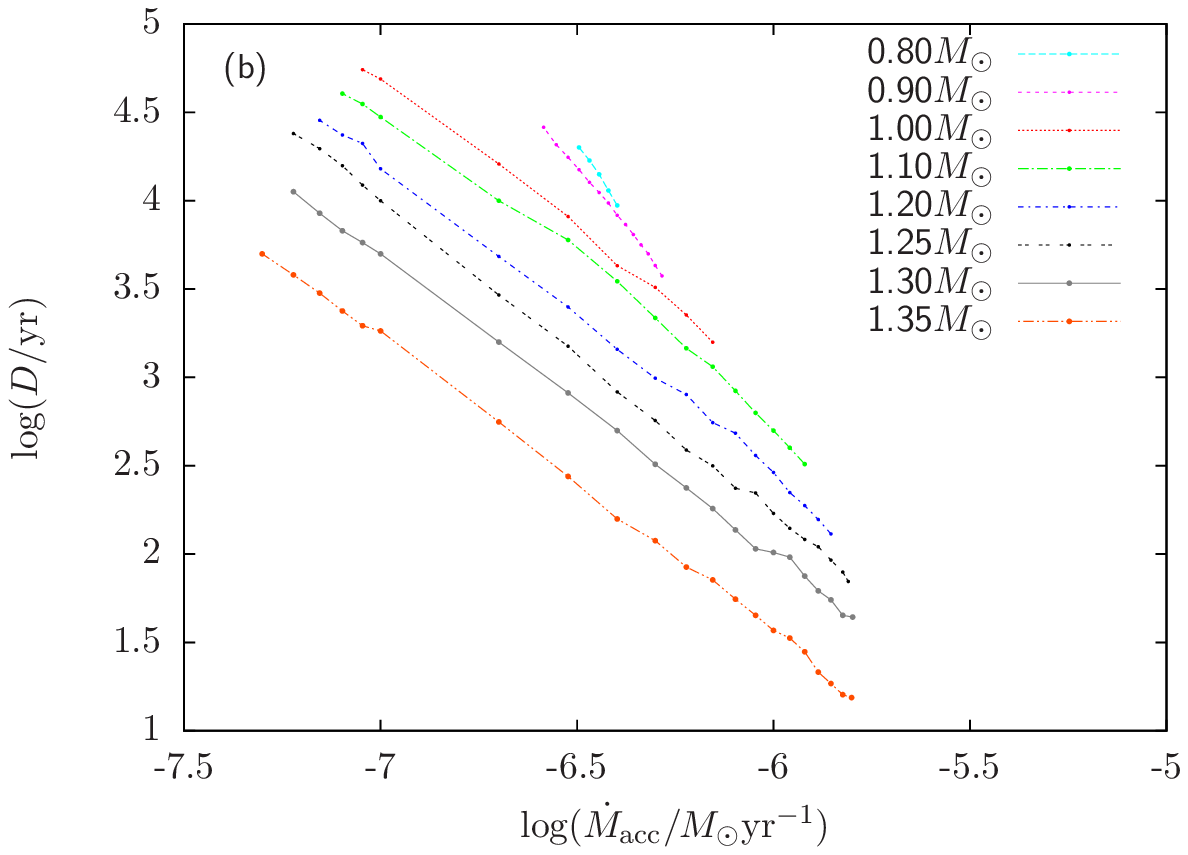}
\caption{Nova cycle durations ($D$) vs. $\dot{M}_{\rm acc}$ for various ${M}_{\rm WD}$. 
Panel (a): the results of H-shell flashes in my simulations.  
Panel (b): the results of He-shell flashes based on the studies of Wu et al. (2017).}
\end{center}
\end{figure}

Nova cycle duration ($D$) is defined as the recurrence time interval between two successive outbursts, which
is an important observed property for nova outbursts. 
Fig. 4 presents nova cycle durations during H-/He-shell flashes for different ${M}_{\rm WD}$ and $\dot{M}_{\rm acc}$. 
There exits a strong inverse relationship between $D$ and  $\dot{M}_{\rm acc}$  for each value of ${M}_{\rm WD}$ (see also Hillman et al. 2016).
For a given ${M}_{\rm WD}$, $D$ become shorter when $\dot{M}_{\rm acc}$ increases.  
This is because nova outbursts  occur when the accumulated  mass of the shell reaches almost the same critical value
for  a specific value of  ${M}_{\rm WD}$ though $\dot{M}_{\rm acc}$ has some effect on the accumulated  mass, 
which means that a higher $\dot{M}_{\rm acc}$ results in a shorter $D$.
Moreover, for a given $\dot{M}_{\rm acc}$ the durations become shorter for massive WDs. 
This is because the shell mass needed for nuclear burning is smaller for massive WDs due to 
their stronger surface gravity. 
Therefore, the recurrent flashes on the massive WDs with higher $\dot{M}_{\rm acc}$ would 
happen more frequently than that of low-mass WDs with lower  $\dot{M}_{\rm acc}$. 
Additionally, the He-nova cycle duration is longer than that of H-nova for a given ${M}_{\rm WD}$ and $\dot{M}_{\rm acc}$. 
This is because the temperature for He burning is higher than that of H burning; it needs 
a thick He-shell for  ignition and thus more time to accrete material.

\section{The single-degenerate model}

In this model, a WD accretes H-/He-rich material from a non-degenerate star that
could be a main-sequence or a
slightly evolved  subgiant star (the WD+MS channel), or a red-giant star
(the  WD+RG channel), or even a He star (the  WD+He star channel).
When the WD grows in mass close to  ${M}_{\rm Ch}$, it may produce an SN Ia 
(see, e.g., Hachisu et al. 1996; Li \& van den Heuvel 1997;  Yungelson \& Livio 1998, 2000;
Langer et al. 2000; Han \& Podsiadlowski 2004).
This model may explain the similarities of most SNe Ia as the WD in this model has the same explosion mass (i.e., ${M}_{\rm Ch}$).
Meanwhile,   there are many SD progenitor candidates of SNe Ia in the observations 
(for more discussions see Sects 3.1.2, 3.2.2 and 3.3.2).

Importantly, this model is supported by some recent observations. 
For example, 
the signatures of  circumstellar matter  (CSM) before SN explosion (e.g., Patat et al. 2007; Wang et al. 2009c; Sternberg et al. 2011; Dilday et al. 2012; 
Silverman et al. 2013a,b),
the early optical and UV emission from ejecta-companion interaction in some SNe Ia (e.g., Kasen 2010; Hayden et al. 2010; Ganeshalingam et al. 2011; 
Wang et al. 2012; Cao et al. 2015; Liu et al. 2015a; Marion et al. 2016; 
but see also Bianco et al. 2011; Shappee et al. 2016; Kromer et al. 2016;  Piro \& Morozova 2016),
the wind-blown cavity in some SN remnants (e.g., Badenes et al. 2007; Williams et al. 2011),  
and the possible pre-explosion images (e.g.,  Voss \& Nelemans 2008; 
McCully et al. 2014), etc. 
It is worth noting that while there is some evidence for gas outflows before SN explosion, this is only seen in a handful of SNe Ia (Ia-CSM). 
It is still unclear what fraction of all SNe Ia have evidence of CSM around them.
Meanwhile,  the pre-explosion images in SN 2012Z and  the UV emission  in iPTF14atg  
mainly relate to type Iax SNe but not normal SNe Ia (e.g., McCully et al. 2014; Cao et al. 2015).

The mass donor in the SD model would survive after SN explosion and potentially be identified, whereas
an SN explosion following the merger of two WDs would leave no compact remnant in the DD model.
Thus, it is a possible way to identify the SD model and the DD model by searching for the surviving companion stars.
It has been suggested that Tycho G may be a surviving companion star of Tycho's SN   
(e.g., Ruiz-Lapuente et al. 2004; for more discussions see Sect. 3.1.3). 
The surviving companion stars predicted by the WD+RG channel may relate to the formation of 
the observed single low-mass He WDs (for more discussions see Sect. 3.2.3), 
and the  surviving companion stars from the WD+He star channel
relate to the formation of  hypervelocity He stars  (for more discussions see Sect. 3.3.3).
Note that Vennes et al. (2017)  recently 
reported the discovery of a low-mass WD (LP 40-365)  with a high proper motion, which 
travels with a velocity greater than the escape velocity of our Galaxy. 
Vennes et al. (2017) found that  LP 40-365 has a peculiar atmosphere that is dominated by intermediate-mass elements,
and argued that  this partially burnt remnant may be ejected by an SN Ia that originates from the SD model. 

The optically thick wind assumption  (see Hachisu et al. 1996) 
is widely adopted in the studies of the SD model (e.g., Li \& van den Heuvel 1997;
Hachisu et al. 1999a,b; Han \& Podsiadlowski 2004; 
Chen \& Li 2007; Meng et al. 2009; Wang et al. 2009a, 2010).
In this assumption,
the red-giant-like regime in Fig. 1 can be replaced by the optically thick wind regime.
If $\dot{M}_{\rm acc}$ exceeds a  critical rate (i.e., $\dot{M}_{\rm cr}$ in Sect. 2), it is supposed that 
the accreted material burns steadily on the surface of the WD at this critical rate; the unprocessed 
material is blown away in the form of the optically thick wind. 
The optically thick wind assumption can enlarge the parameter space for producing SNe Ia and thus their rate 
(e.g., Li \& van den Heuvel 1997;  Han \& Podsiadlowski 2004).
Meanwhile, the properties of some supersoft X-ray sources and recurrent novae may be explained by this assumption
 (e.g., Hachisu \& Kato 2003, 2005, 2006; Hachisu et al. 2007; Kato et al. 2018).

However, the optically thick wind assumption is still under hot debate. 
For example,
the metallicity  threshold predicted by this model is conflict with observations
(e.g., Prieto et al. 2008; Badenes et al. 2009a; Galbany et al. 2016).
Meanwhile,
SNe Ia are not expected at high-redshift ($z>1.4$) for this model (see Kobayashi et al. 1998),
but some high-redshift SNe Ia even at $z=2.26$ have been reported 
(e.g., Graur et al. 2011, 2014a; Frederiksen et al. 2012; Rodney et al. 2012, 2014, 2015; Jones et al. 2013).
In addition,
the wind velocity predicted by this model is too large to match observations (e.g., Patat et al. 2007; Badenes et al. 2007).
According to this model,  the hot WD would photoionize the surrounding interstellar medium, 
so some emission lines (such as He II 4686 and [O I] 6300) can be produced.
And, an emission-line shell or nebula should be visible around the progenitor up to thousands of years after the explosion. 
These emission lines should be detected in old elliptical galaxies or around some individual SN remnants.
However, any evidence for such emission has not been found so far 
(e.g., Woods \& Gilfanov 2013, 2016; Graur et al. 2014b; Johansson et al. 2014, 2016; Woods et al. 2017).
Note that  some alternative models to the optically thick wind assumption have been proposed, 
for example, the super-Eddington wind model (e.g., Ma et al. 2013; Wang et al. 2015a) 
and the common-envelope wind model (see Meng \&  Podsiadlowski 2017).

Additionally, a serious challenge to the SD model is the non-detection of stripped H-rich material.
In the SD model, H-rich material can be removed from the surface of the non-degenerate companion star.
Recent 3D hydrodynamic simulations of the interaction between SN Ia ejecta and their MS/RG companion stars
indicate that the stripped H-rich material is always larger than 0.1$\,{M}_\odot$ (see Pan et al. 2012, 2014; Liu et al. 2012a, 2013a).
However, no stripped H-rich material has been detected in late-time spectra of  SNe Ia yet 
(e.g., Leonard et al. 2007; Shappee et al. 2013; Lundqivst et al. 2013, 2015; Maguire et al. 2015).\footnote{Late-time observations  
can afford a new diagnostic of SN Ia nebular, explosion, and progenitor physics. Graur et al. (2017a) recently summarized 
the progress in this field.  The relevant theoretical investigations on this field  include Fransson \& Kozma (1993), Seitenzahl et al. (2009) and R\"{o}pke et al. (2012). 
Meanwhile, the recent observational studies include Fransson \& Jerkstrand (2015), Graur et al. (2016, 2017a), Shappee et al. (2017), 
Kerzendorf et al. (2017a), Dimitriades et al. (2017) and Yang et al. (2018).}
Furthermore, the SD model is suffering the issue of the deficit of the supersoft X-ray flux in the observations
(e.g., Gilfanov \& Bogd$\acute{\rm a}$n 2010; Di Stefano 2010). Note that the supersoft X-ray source stage 
only accounts for a short time in the SD model, which can be alleviating the existing X-ray constraints (see also Wang \& Han 2012).

\subsection{The WD+MS channel}
This channel is usually called the supersoft channel, in which  a CO
WD accretes H-rich material from a MS or a
slightly evolved subgiant star. The accreted H-rich material
is burned into He, and then the He is converted to carbon and
oxygen. The WD  may explode as an SN~Ia  when it grows in mass close to ${M}_{\rm Ch}$.
For more discussions on  this channel, see, e.g.,  Li \& van den Heuvel 1997;
Hachisu et al. (1999a, 2008),  Langer et al. (2000), 
Han \& Podsiadlowski (2004), Fedorova et al. (2004), 
Meng et al. (2009), Wang et al. (2010, 2014a), Chen et al. (2014a),
Meng \& Podsiadlowski (2017) and Liu \& Stancliffe (2017, 2018).

\subsubsection{Evolutionary scenarios and  parameter space}

\begin{figure}
\begin{center}
\epsfig{file=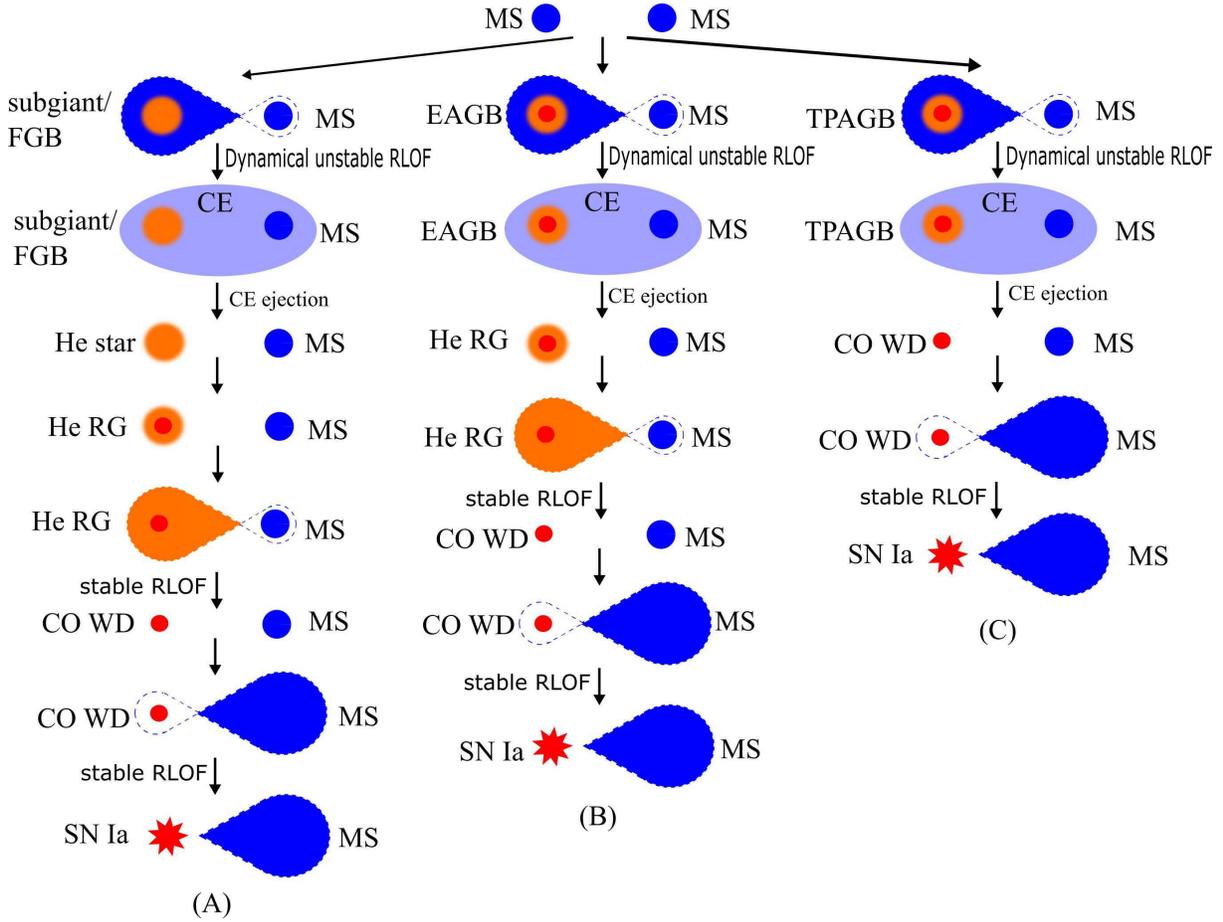,width=16cm} \caption{Evolutionary scenarios to WD+MS systems that can form SNe Ia (see also Wang \& Han 2012).}
\end{center}
\end{figure}

In the supersoft channel, SNe Ia originate from the evolution of WD+MS systems.
Fig. 5 presents the main binary evolutionary scenarios to WD+MS systems  that can form SNe Ia    
(for details see Wang \& Han 2012; see also Meng et al. 2009; Wang et al. 2010).
For Scenario A, the initial parameters of the primordial binaries are in the range of
$M_{\rm 1,i}\sim4.0$$-$$7.0\,M_\odot$, $q=M_{\rm 2,i}/M_{\rm 1,i}\sim0.3-0.4$, and $\log P^{\rm i}\rm(d) \sim 1.0-1.5$, 
in which $M_{\rm 1,i}$ and $M_{\rm 2,i}$  are the initial masses of the primordial primary and secondary, 
respectively, and $P^{\rm i}$ is the initial orbital period of the primordial systems. 
For Scenario B, the initial binary parameters are in the range of
$M_{\rm 1,i}\sim2.5$$-$$6.5\,M_\odot$, $q=M_{\rm 2,i}/M_{\rm 1,i}\sim0.2-0.9$, and $\log P^{\rm i}\rm(d) \sim 2.0-3.0$.
For Scenario C, the initial binary  parameters  are in the range of
$M_{\rm 1,i}\sim3.0$$-$$6.5\,M_\odot$, $q=M_{\rm 2,i}/M_{\rm 1,i}\sim0.2-0.7$, and $\log P^{\rm i}\rm(d) \sim 2.5-3.5$.
Among the three scenarios, SNe Ia are mainly produced by Scenarios A and B, in which each scenario 
contributes to about half of SNe Ia through the supersoft channel (see Wang et al. 2010).

\begin{figure}
\begin{center}
\epsfig{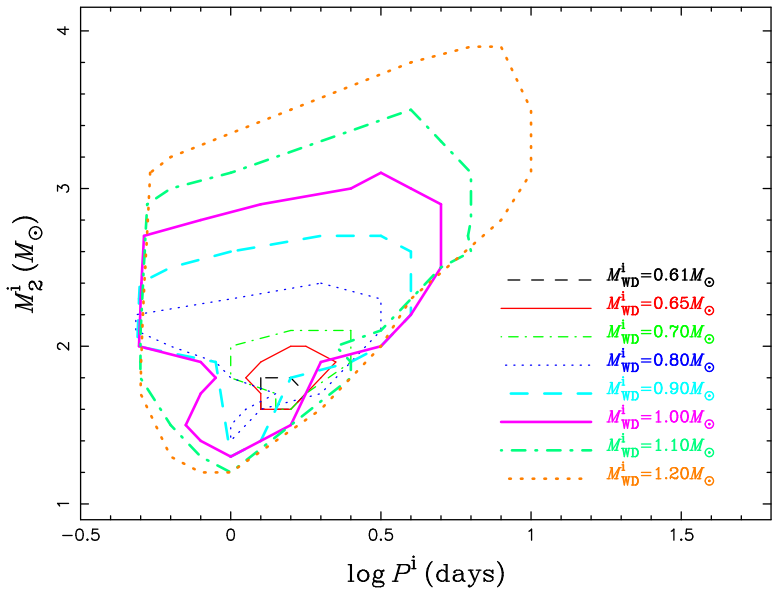} \caption{Initial parameter space of SNe Ia for the supersoft channel
in the $\log P^{\rm i}-M^{\rm i}_2$ plane with different ${M}^{\rm i}_{\rm WD}$. Source: From Wang et al. (2010).}
\end{center}
\end{figure}

After the formation of WD+MS systems, the WD can accrete material from a MS or a slightly evolved subgiant star. 
According to the optically thick wind assumption, 
Li \& van den Heuvel (1997) studied the supersoft channel based on detailed binary evolution computations with ${M}^{\rm i}_{\rm WD}=1.0$, 1.2$\,M_\odot$. 
Han \& Podsiadlowski (2004)  investigated   this channel in a systematic way with different ${M}^{\rm i}_{\rm WD}$ and gave the results of BPS approach. 
Fig. 6 shows the initial parameter space of SNe Ia for the supersoft channel in the $\log P^{\rm i}-M^{\rm i}_2$ plane with different ${M}^{\rm i}_{\rm WD}$,
where $M^{\rm i}_2$ is the initial mass of the MS star and $P^{\rm i}$ is the initial orbital period  of the WD+MS system. 
If  the initial parameters of a WD+MS system are 
located in this parameter space, an SN Ia  is supposed to be formed.
The minimum ${M}^{\rm i}_{\rm WD}$ for producing SNe Ia in this channel is $0.61\,M_\odot$ that can 
grow in mass to ${M}_{\rm Ch}$.  
According to a detailed BPS
simulation, Wang et al. (2010) estimated that the Galactic SN Ia rate from the  supersoft channel is $\sim$$1.8\times10^{-3}\,{\rm yr}^{-1}$,
mainly contributing to the observed SNe Ia with intermediate and long delay times.
Note that if the new mass-accumulation efficiencies ($\eta$) in Fig. 2 are adopted, some systems 
with low mass MS donors in Fig. 6 will not produce SNe Ia as the values of the new $\eta$
are lower than those used in Wang et al. (2010).

SN 2002ic is a peculiar SN Ia that lost a few solar mass of H-rich material before SN explosion (e.g., Hamuy et al. 2003; Deng et al. 2004).
Han \& Podsiadlowski (2006) suggested that the 
atypical properties of SN 2002ic may be reproduced by the delayed dynamical instability in the frame of the supersoft channel,
which requires that the mass donor was initially $\sim$3\,$M_{\odot}$.
Han \& Podsiadlowski (2006) estimated that $<$1\% SNe~Ia should belong to 
SN 2002ic-like objects (see also Meng et al. 2009).

\subsubsection{Progenitor candidates}

In the observations, the candidates of the supersoft channel have been identified as  supersoft X-ray sources and recurrent novae
(e.g., van den Heuvel et al. 1992; Rappaport et al. 1994).
Supersoft X-ray sources  are strong candidates of SN Ia progenitors, 
which are WD binaries where steady nuclear  burning occurs on the surface of the WDs (e.g., Chen et al. 2015).
Recurrent novae usually include a massive WD with  $\dot{M}_{\rm acc}<\dot{M}_{\rm st}$.
Especially,
U Sco (a recurrent nova) is a strong progenitor candidate of SNe Ia, 
including a $1.55\pm0.24\,M_{\odot}$ WD  and a $0.88\pm0.17\,M_{\odot}$ MS donor
with an orbit period of $\sim0.163$\,d (e.g., Hachisu et al. 2000; Thoroughgood et al. 2001).
However, Mason (2011) argued that U Sco may be a nova outburst happened on the surface of an ONe WD, and  thus its
final fate may not be an SN~Ia  but collapse to a neutron star.

In addition, M31N~2008-12a is a remarkable recurrent nova in M31, and its recurrence period is $\sim$1\,yr;
the WD mass in M31N~2008-12a may be  
$\sim$$1.38\,M_{\odot}$ with $\dot{M}_{\rm acc}=1.6\times10^{-7}\,{M}_\odot\,\mbox{yr}^{-1}$, making it
a promising candidate of SN Ia progenitors (e.g., Darnley et al. 2014, 2016; Tang et al. 2014; Kato 
et al. 2015, 2017).
In order to search the progenitor candidates of SNe Ia, 
Rebassa-Mansergas et al. (2017)  recently obtained a large sample of detached WD+F/G/K star binaries (see also Li et al. 2014),
and Toonen et al. (2017) made a detailed estimate of the number of WD+MS binaries in the Gaia sample.




\subsubsection{Surviving companion stars}
According to the supersoft channel, Han (2008) gave various properties of 
the surviving companion stars at the moment of
SN explosion, which
are runaway stars that are moving away from the center of SN remnants (see also Wang \& Han 2010a).  
The surviving companion star in the supersoft channel would evolve to a CO WD
finally. Hansen (2003) argued that the supersoft channel might
potentially explain the properties of high-velocity WDs in the halo, 
which differs from others as they consist exclusively of single stars.
In order to 
search the surviving companion stars after  SN explosion,
Pan et al. (2012) carried out  the impact of SN Ia
ejecta on MS, RG and He star companions based on hydrodynamical simulations
(see also Pan et al. 2010, 2013, 2014; Liu et al. 2012a, 2013a,b).

It has been suggested that Tycho G may be a surviving companion star of Tycho's SN, 
which has a space velocity of $136\,{\rm km/s}$ (see Ruiz-Lapuente et al. 2004). However,
the surviving companion star of Tycho's SN is still not well determined. 
Han (2008) found that the observed properties of Tycho G are compatible with 
the surviving companion star of the supersoft channel, e.g., 
the surface gravity, the effective temperature  and the space velocity, etc (see also Wang \& Han 2010a).
Lu et al. (2011) claimed that the non-thermal X-ray arc in Tycho's SN remnant  
may originate from  the interaction between SN ejecta and the stripped mass of the companion.
In addition, Zhou et al. (2016) suggested that 
the most plausible origin for the expanding  molecular bubble surrounding Tycho's SN remnant 
is the fast outflow driven from a WD as it accreted material from a non-degenerate donor,
which provides an evidence for a SD progenitor for Tycho's SN.
Note that Fang et al. (2018) recently argued that  the SN ejecta evolved in the cavity driven by the latitude-dependent wind provides
an alternative explanation for the peculiar shape of the periphery of Tycho's SN remnant.
For more studies on the surviving companion star of Tycho'SN, see, e.g., Fuhrmann (2005), Ihara et al.
(2007), Gonz$\acute{\rm a}$lez-Hern$\acute{\rm a}$ndez et al. (2009), Kerzendorf et al. (2009, 2013), 
 Liu et al. (2013a) and Pan et al. (2014). 
Note that it is still no conclusive confirmation about 
any surviving companion stars of SNe~Ia.

\subsection{The WD+RG channel}
This channel is called the symbiotic channel, usually consisting of
a hot WD and a RG star. In most cases a hot WD accretes material from a RG star
through stellar wind, but in some cases through Roche-lobe. 
The surviving companion stars in this channel may relate to the formation of the observed 
single low-mass He WDs (for more discussions see Sect. 3.2.3).
For more discussions on this channel see, e.g.,  Yungelson et al. (1995), Hachisu et al. (1996, 1999b),  
Li \& van den Heuvel (1997), Yungelson \& Livio (1998), L\"{u} et al. (2006, 2009),
Xu \& Li (2009), Wang et al. (2010), Wang \& Han  (2010b), 
Chen et al. (2011) and Liu et al. (2018a).

\subsubsection{Evolutionary scenario and  parameter space}

\begin{figure}
\begin{center}
\epsfig{file=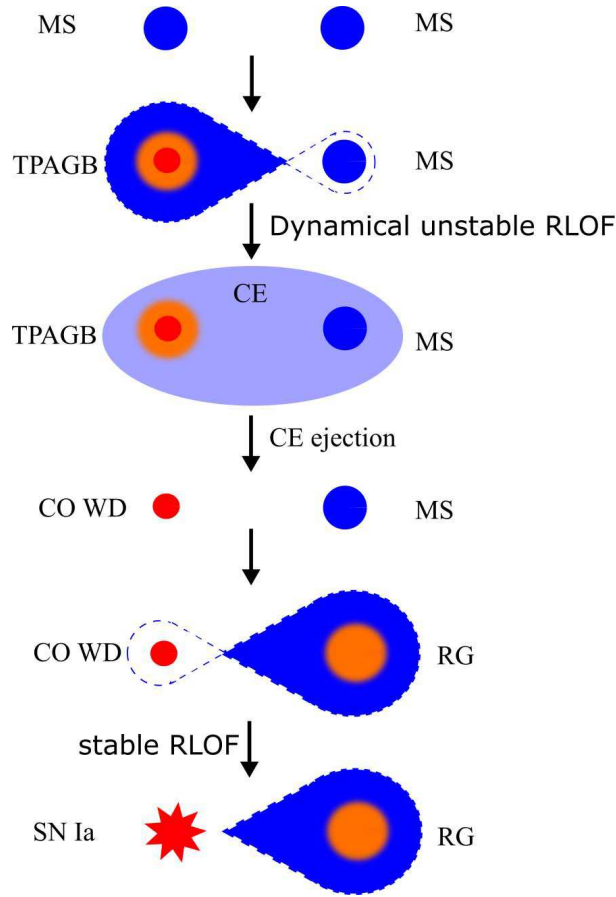,width=8cm} \caption{Evolutionary scenario to WD+RG systems that can form SNe Ia  (see also Wang \& Han 2012).}
\end{center}
\end{figure}

Compared with the supersoft  channel,
SNe Ia in the symbiotic channel originate from wider primordial binaries.
Fig. 7 shows the binary evolutionary scenario to WD+RG systems  that can form SNe Ia (for details see Wang \& Han 2012; see also Wang et al. 2010).
There is one binary evolutionary scenario that can
produce CO WD+RG systems and then form SNe~Ia. 
The primordial primary first fills its Roche-lobe when it evolves to the thermal pulsing asymptotic giant branch (TPAGB) stage. 
A common-envelope  (CE) may be formed owing to the dynamically unstable Roche-lobe overflow (RLOF). The primordial primary becomes a CO
WD after the CE ejection. At this moment, a CO WD+MS system is formed.
A CO WD+RG system can be formed when the MS companion evolves to the RG stage.
For the symbiotic channel, SN Ia explosions
happen for the binary parameter ranges of
$M_{\rm 1,i}\sim5.0$$-$$6.5\,M_\odot$, $q=M_{\rm 2,i}/M_{\rm 1,i}\sim0.1-0.5$, 
and $\log P^{\rm i}\rm(d) \sim 2.5-4.0$  (see Liu et al. 2018a).

Previous studies suggested that 
the initial parameter space for producing SNe Ia from the symbiotic channel is too
small as  a CE  is easily formed  when the RG star fills its Roche lobe, and thus 
a low rate of SNe Ia (e.g., Yungelson et al. 1995; Li \& van den Heuvel 1997,
Yungelson \& Livio 1998; L\"{u} et al. 2006; Wang et al. 2010).  
In order to  avoid
the formation of  a CE once the RG star fills its Roche-lobe, Hachisu et al. (1999b) supposed that 
a stellar wind from the WD strips some mass of
the RG star to stabilize the mass-transfer process, known as the mass-stripping model.
However, this model has not  been confirmed by the observations.
Note that some studies  enlarged the initial parameter space for producing 
SNe Ia  and thus obtained a high rate through the symbiotic channel
(e.g., L\"{u} et al. 2006; Chen et al. 2011) ,
but these works strongly depends on model parameters or assumptions.

\begin{figure}
\begin{center}
\epsfig{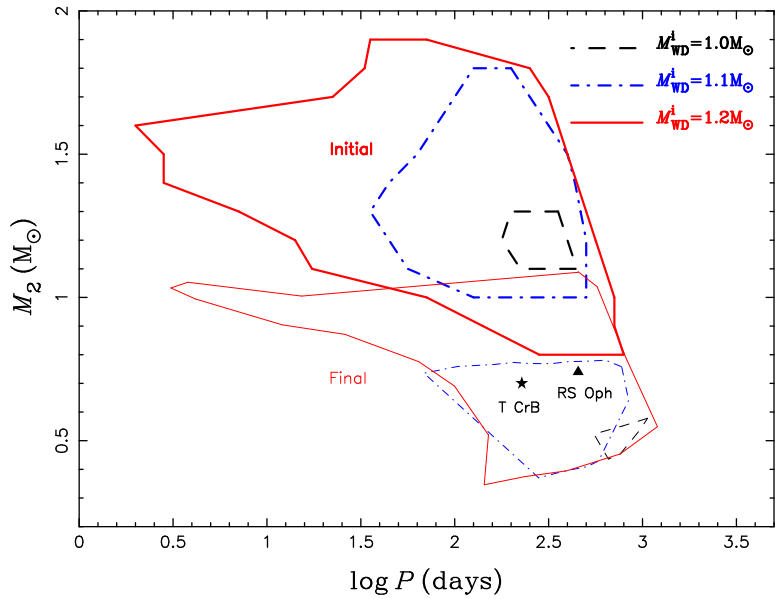} \caption{Initial and final parameter space of SNe Ia for the symbiotic channel
in the $\log P-M_2$ plane with different ${M}^{\rm i}_{\rm WD}$.  
The filled asterisk and triangle  show the locations of symbiotics T CrB 
and RS Oph, respectively.   
The data points of these contours  are from Liu et al. (2018a).}
\end{center}
\end{figure}

Liu et al. (2018a) recently  adopted an integrated mass-transfer prescription for the symbiotic channel based on a power-law adiabatic supposition, 
which is applicable for the mass-transfer from a RG star onto the WD (see Ge et al. 2010).
They evolved a large number of WD+RG systems, and found  that the parameter space of WD+RG systems for producing SNe Ia  is significantly enlarged. 
The mass-transfer prescription adopted by Liu et al. (2018a)  is still under debate when the RG star fills its Roche-lobe 
(see Woods \& Ivanova 2011), 
but their work at least gave an upper limit of the parameter space  for producing SNe Ia.
Fig. 8 shows the initial and final parameter space of SNe Ia in the $\log P-M_2$ 
plane with different ${M}^{\rm i}_{\rm WD}$ for the symbiotic channel. 
The minimum ${M}^{\rm i}_{\rm WD}$ for producing SNe Ia
 in this channel  is $\sim$$1.0\,M_\odot$. 
The binary parameters of  RS Oph and T CrB are located in the parameter space of WD+RG systems for producing SNe Ia (see Fig. 8);
these two symbiotics may form SNe Ia in their future evolutions (for more discussions see Sect. 3.2.2). 
According to a detailed BPS approach, Liu et al. (2018a) found that the symbiotic channel may contribute to at most 2\% of all SNe Ia in our Galaxy, 
and mainly contribute to SNe Ia with intermediate and long delay times. 
The rate of SNe Ia in Liu et al. (2018a) is 
still low compared with previous studies as  
most of WD+RG systems are difficult to locate in the initial parameter space of Fig. 8
in the current BPS studies, which needs to be further investigated.

\subsubsection{Progenitor candidates}

Symbiotic novae have been proposed as progenitor candidates of SNe Ia, which are binaries where
the WD accretor undergoes a classical nova eruption. In the observations,
many symbiotic novae have WD mass 
close to ${M}_{\rm Ch}$  and have giant companions, e.g., RS Oph,  T CrB and V745 Sco, etc
(e.g., Hachisu \& Kato 2001; Parthasarathy et al. 2007).
(1)
RS Oph has a 1.2$-$$1.4\,M_\odot$ WD and a 0.68$-$$0.8\,M_\odot$ RG star with an orbital period of $\sim$454\,d  (e.g., Brandi et al. 2009).
Miko\l{}ajewska \& Shara (2017) recently suggested that the WD in RS Oph may be a CO WD by analyzing its spectra, making it a  strong progenitor candidate of SNe Ia. 
(2) 
T CrB has a $\sim$$1.2\,M_\odot$  WD and a $\sim$$0.7\,M_\odot$ RG star with an orbital period of $\sim$227\,d 
 (e.g., Belczy$\acute{\rm n}$ski \& Miko\l{}ajewska 1998).
However, it is still uncertain 
whether the WD in T CrB is a CO WD or an ONe WD;
the latter is expected to result in accretion-induced collapse rather than an SN Ia explosion. 
(3) 
V745 Sco is a symbiotic nova. 
Orlando et al. (2017) recently suggested that the WD in V745 Sco is a CO WD as 
this nova shows no signs of Ne enhancement.
Furthermore, the ejected mass during nova outbursts in V745 Sco is considerably 
lower than the mass needed to initiate the thermonuclear reaction (e.g., Drake et al. 2016),  making it a strong progenitor candidate of SNe Ia.

Tang et al. (2012) recently found a peculiar symbiotic J0757, including a 
$1.1\pm0.3\,M_\odot$ WD and a $0.6\pm0.2\,M_\odot$ RG star with an orbit period of $\sim$119\,d. 
J0757 does not show any signature of  symbiotic stars in quiescent stage, which
is different from any other known classic or symbiotic novae.
This implies that it is a missing population among symbiotics.   In addition,
J0757 had a 10\,yr flare in the 1940s, possibly from H-shell
burning on the surface of the WD and without significant
mass-loss, indicating that the WD in J0757 could increase mass effectively and may explode as an SNe Ia in the future.
It is worth noting that the rate of symbiotic
novae can put some constraints on the formation of SNe Ia, and thus more symbiotic
novae are needed in the observations.

\subsubsection{Surviving companion stars}
The surviving companion stars of SNe Ia from the symbiotic channel relate to the
formation of single low-mass He WDs (LMWDs; $<$$0.45\,M_{\odot}$), 
the existence of which is supported by some observations (e.g., Marsh et al. 1995; Kilic
et al. 2007).  Kalirai et al. (2007) 
suggested that single stars may form single LMWDs,   especially at high metallicity environment (see also Kilic et al. 2007).
However, the study of the initial-final mass
relation for stars with different metallicities  indicated that only LMWDs with
mass $>$$0.4\,M_{\odot}$ can be formed through this way (e.g., Han et al. 1994; Meng et al. 2008).

Single LMWDs can be naturally produced in binaries, in which their compact
companions exploded as SNe~Ia. 
The surviving companion stars of the old SNe~Ia from the  symbiotic channel have low masses ($<$$0.45\,M_{\odot}$), 
the final fate of which is single LMWDs (e.g., Justham et al. 2009; Wang \& Han 2010a). 
On the other hand, the existence of 
the single LMWDs indicates that  some SNe~Ia may be happened with RG donors in symbiotics. 
Note that
Nelemans \& Tauris (1998) argued that single LMWDs might be formed
through a solar-like star accompanied by a brown dwarf or a massive planet with
a relatively close orbit.  Note also that
Zhang et al. (2018) recently claimed that  the merger remnants of He WD+MS systems can provide an alternative way for the formation of
single LMWDs.

\subsection{The WD+He star channel}

The mass donor in this channel is  a He star or a He subgiant, 
which can afford enough mass  for the WD growing in mass to ${M}_{\rm Ch}$ and forming an SN Ia finally.
This channel is  known as the He star donor channel, which
is a particularly favorable way for producing observed young SNe Ia (see Wang et al. 2009a,b).
The surviving companion stars in this channel may be related to the formation of 
hypervelocity He stars  (for more discussions see Sect. 3.3.3).
For more discussions on this channel, see, e.g., Yoon \& Langer (2003),  
Ruiter et al. (2009), Wang \& Han (2010c), Liu et al. (2010),  Claeys et al. (2014), Wu et al. (2016) and Wang et al. (2017a).

\subsubsection{Evolutionary scenarios and  parameter space}

\begin{figure}
\begin{center}
\epsfig{file=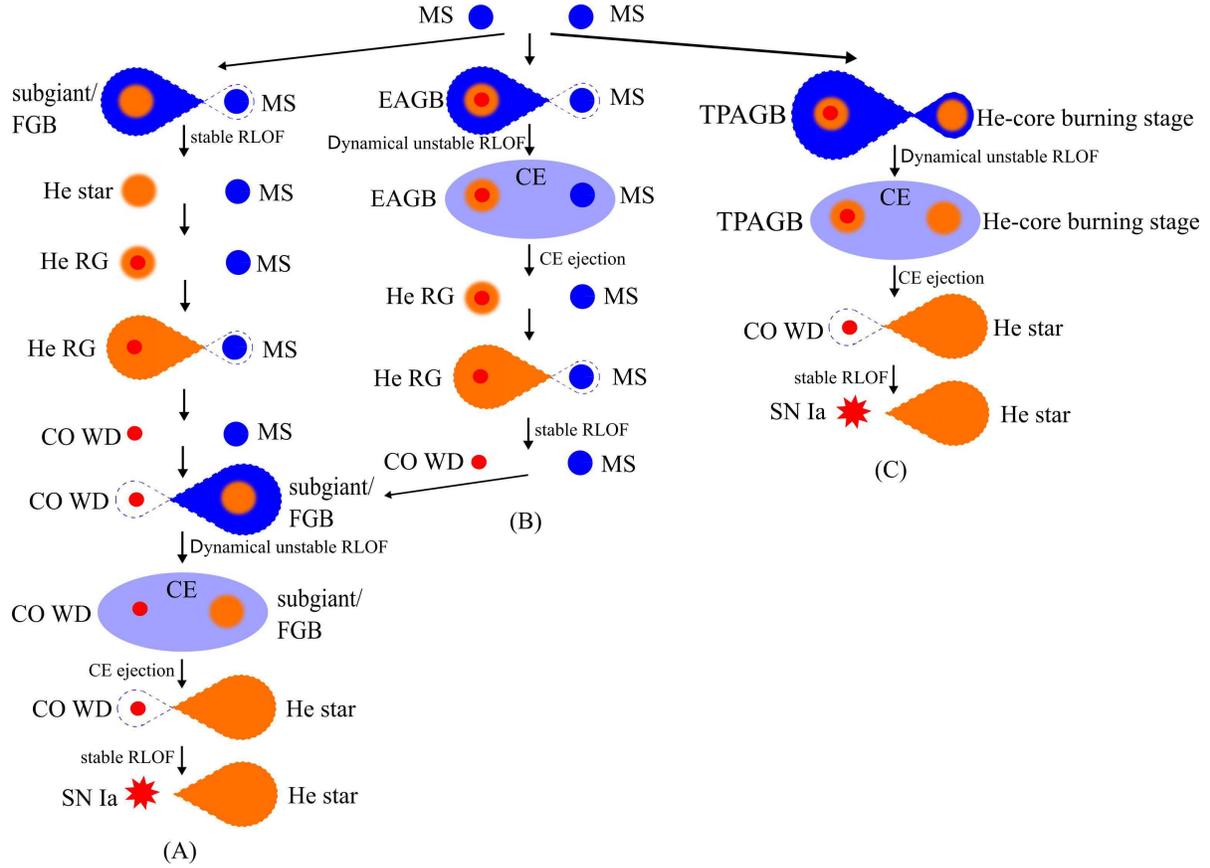,width=16cm} \caption{Evolutionary scenarios to WD+He star systems that can form SNe Ia  (see also Wang \& Han 2012).}
\end{center}
\end{figure}
 
Fig. 9 shows the  binary evolutionary scenarios to WD+He star systems that can form SNe Ia    
(for details see Wang \& Han 2012; see also Wang et al. 2009b).
Three  evolutionary
scenarios can produce WD+He star systems and then form SNe~Ia.
For Scenario A, the initial parameters of the primordial binaries are in the range of
$M_{\rm 1,i}\sim5.0$$-$$8.0\,M_\odot$, $q=M_{\rm 2,i}/M_{\rm 1,i}\sim0.2-0.9$, and $\log P^{\rm i}\rm(d) \sim 1.0-1.5$.
For Scenario B, the initial binary parameters are in the range of
$M_{\rm 1,i}\sim6.0$$-$$6.5\,M_\odot$, $q=M_{\rm 2,i}/M_{\rm 1,i}>0.9$, and $\log P^{\rm i}\rm(d) \sim 2.5-3.0$.
For Scenario C, the initial binary  parameters  are in the range of
$M_{\rm 1,i}\sim5.0$$-$$6.5\,M_\odot$, $q=M_{\rm 2,i}/M_{\rm 1,i}>0.9$, and $\log P^{\rm i}\rm(d)>3.0$.
Among the three scenarios, Scenario A contributes to almost 90\% SNe Ia through the  He star donor channel (see Wang et al. 2009b).

\begin{figure}
\begin{center}
\epsfig{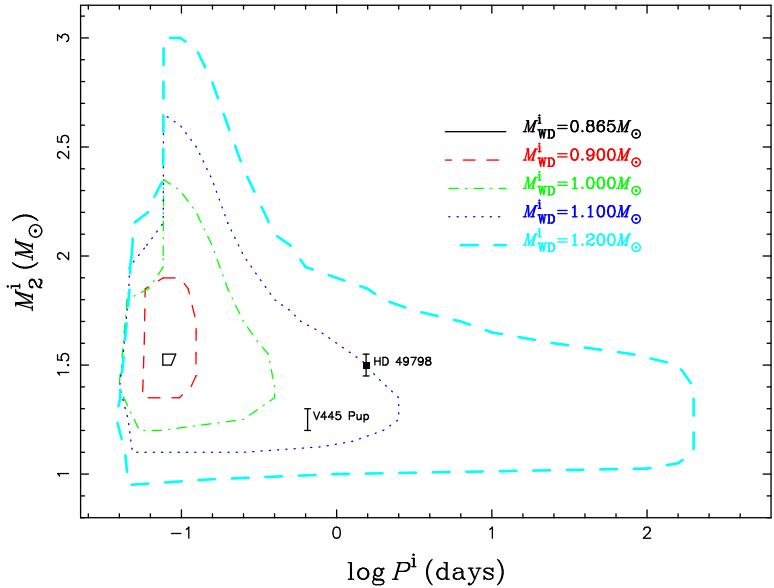} \caption{Initial parameter space of SNe Ia for the He star donor channel. 
The locations of V445 Pup
and HD 49798 with its WD companion  are indicated in this figure. 
The data points of these contours  are from  Wang et al. (2009a).}
\end{center}
\end{figure}

Adopting the optically thick
wind assumption, Wang et al. (2009a) investigated  the He star donor
channel in a systematic way, in which they performed binary evolution
computations for about 2600 close WD+He star systems.  They
determined the initial parameter space of WD+He star systems that can lead to SNe Ia in the $\log P^{\rm i}-M^{\rm i}_2$ plane (see Fig. 10). 
The minimum ${M}^{\rm i}_{\rm WD}$ for producing SNe Ia  in this channel could be
as low as $0.865\,M_{\odot}$.  The binary parameters of  V445 Pup and HD 49798
with its WD companion are located in the parameter space of WD+He star systems for producing SNe Ia (see Fig. 10), which means that they are
progenitor candidates of SNe Ia (for more discussions see Sect. 3.3.2).
The Galactic SN~Ia rate from this channel is
$\sim$$0.3\times 10^{-3}\,{\rm yr}^{-1}$ and  this channel can
produce the observed SNe~Ia with short delay times ($\sim$45$-$140\,Myr; see Wang et al. 2009b). Wang \& Han (2010c) suggested that
SNe Ia from the He star donor channel occur systemically later in low-metallicity environments.
By considering the possibility of the off-center carbon burning,  Wang et al. (2017a) estimated that
the Galactic SN Ia rates from the He star donor channel decrease to $\sim$$0.2\times 10^{-3}\,{\rm yr}^{-1}$ based on a detailed BPS 
method.

\subsubsection{Progenitor candidates}

For the He star donor channel, two massive WD+He star systems are good candidates of SN~Ia progenitors, 
i.e., V445 Pup and HD 49798 with its WD companion.

V445 Pup is the only He nova discovered so far, which was detected during its outburst in late 2000 
(e.g., Ashok \& Banerjee 2003; Kato \& Hachisu 2003). According to the light curve fitting of V445 Pup, 
 Kato et al. (2008) suggested that the WD have mass $\gtrsim$$1.35\,M_{\odot}$  and
 half of the accreted material  still remains on its surface. 
Woudt et al. (2009) obtained the mass of the He star donor $\sim$1.2$-$$1.3\,M_{\odot}$ based on  the pre-outburst luminosity of the
binary (see also Piersanti et al. 2014).
Goranskij et al. (2010)  suggested that
the most probable orbital period for this binary is $\sim$0.65\,d.
The binary parameters of V445 Pup are located in the parameter space contours for producing SNe Ia (see Fig. 10).
In addition, Woudt
\& Steeghs (2005) suggested that
the WD in V445 Pup is a CO WD but not an ONe WD as no signatures of Ne enhancement were detected. 
Thus, I speculate that V445 Pup is a strong candidate of SN Ia progenitors.

HD 49798 is a subdwarf O6 star (1.50$\pm$0.05$\,M_{\odot}$), including 
a massive compact companion (1.28$\pm$0.05$\,M_{\odot}$) with an orbital period of
1.548\,d (e.g., Thackeray 1970; Bisscheroux et al. 1997; Israel et al. 1997; Mereghetti et al. 2009). 
However, the nature of the  compact companion is still not well known (e.g., 
Bisscheroux et al. 1997; Liu et al. 2015b; Mereghetti et al. 2016; Popov et al. 2018).
Mereghetti et al. (2016) claimed that the companion of HD 49798 is more likely 
a neutron star based on 
the new angular momentum and magnetic field analysis  (see also Brooks et al. 2017a), but
Popov et al. (2018)  recently stated that the continuous stable spin-up of the 
compact companion  can be reproduced  through contraction of a young WD.
Assuming the companion of HD 49798 is a CO WD, 
Wang \& Han (2010d) suggested that the massive
WD can grow in mass to ${M}_{\rm Ch}$ after about $10^{4}$\,yr based on a detailed
binary evolution computations.   However,  Wang et al. (2017a) recently argued that
off-center carbon burning may occur when the WD increases its mass close to ${M}_{\rm Ch}$ 
owing to a high mass-transfer rate ($>$$2.05\times 10^{-6}\,{M}_\odot\,\mbox{yr}^{-1}$).
Thus,  the WD companion of  HD 49798 may form a neutron star but  not an SN Ia eventually.

\subsubsection{Surviving companion stars}

The surviving companion stars of SNe Ia from the  He star donor channel relate to 
the formation of hypervelocity stars (HVSs), which are stars that can  escape the gravitational pull of the Galaxy.
The first HVS is a B-type star with a Galactic rest-frame radial velocity of 673\,km/s, 
which was discovered serendipitously by Brown et al. (2005). 
Up to now, over 20 HVSs have been confirmed by the observations  (e.g., Hirsch et al. 2005; 
Edelmann et al. 2005;  Li et al. 2012; Zhong et al. 2014;
Zheng et al. 2014;   Brown et al.  2014; Huang et al. 2017).
It has been suggested that HVSs
can be produced by the tidal disruption of a binary through
interaction with the super-massive black hole at the Galactic
center (see, e.g., Hills1988; Yu \& Tremaine 2003; Lu et al. 2010; Zhang et al.
2010). For a recent review on HVSs see Brown (2015).

To date, most of HVSs discovered are B-type stars (see Brown 2015).
Only one HVS (US 708, HVS2)   is an extremely He-rich sdO star
in the Galactic halo (see Hirsch et al. 2005). 
Wang \& Han (2009) studied the properties of the surviving companion stars of 
SNe~Ia from the He star donor channel, and suggested that 
this channel provides an alternative way
for the production of hypervelocity He stars such as US 708 (see also
Justham et al. 2009). Ziegerer et al. (2017) recently found that J2050 is
the spectroscopic twin  of US 708, which
could be surviving companion stars of SNe Ia that happened in WD+He star systems.
Note that Geier et al. (2015)  recently  presented a spectroscopic and kinematic analysis of US 708, 
and found that it  is the currently fastest unbound star in our Galaxy with a velocity of $\sim$1200\,km/s. 
Geier et al. (2015) suggested that the surviving  donors  of sub-${M}_{\rm Ch}$ 
double-detonation SNe Ia (see Sect. 5) may explain such high velocity 
due to the short orbital periods at the moment of SN explosion.
 
In order to identify the surviving companion stars of the He star donor channel, 
Pan et al. (2010)  carried out the impact of the SN explosion on the He donors
based on hydrodynamical
simulations  (see also Pan et al. 2013, 2014; Liu et al. 2013b). It is worth noting that
some ongoing surveys are searching for more hypervelocity He stars that originate from
surviving donors of SNe Ia,  for example,  the LAMOST LEGUE survey (e.g., Deng et
al. 2012) and the Hyper-MUCHFUSS project (e.g., Tillich et al. 2011; Geier et al.  2011, 2015).

\section{The double-degenerate model}

In the classical DD model, SNe Ia result from the merging of double WDs with total mass $\geq$${M}_{\rm Ch}$;
the merging of two WDs is due to the gravitational wave radiation that
drives orbital inspiral to merger  (e.g., Webbink 1984; Iben \& Tutukov 1984).
It has been suggested that this model can reproduce  the observed rates and delay time distributions of SNe Ia (e.g., 
Nelemans et al. 2001; Ruiter et al. 2009, 2013; Mennekens et al. 2010; 
Yungelson \& Kuranov 2017; Liu et al. 2018b), and 
may explain the formation of some observed super-luminous SNe~Ia 
that have WD explosion masses $\geq$$2\,M_\odot$ (e.g., Howell et al. 2006; 
Hichen et al. 2007; Scalzo et al. 2010; Silverman et al. 2011).

One of the strongest pieces of evidence in favor of the DD model is 
the power-law delay time distribution with an index of $-1$ (e.g.,  Maoz \& Mannucci 2012; Maoz \& Graur 2017).
This delay time distribution likely explains correlations between the SN Ia rates and galaxy properties  
(e.g., Graur \& Maoz 2013; Graur et al. 2017b).
The DD model is also supported by some other observational facts.
For example, 
the absence of H and He lines in the nebular spectra of most SNe Ia
(e.g., Leonard 2007; Ganeshalingam et al. 2011; Maguire et al. 2015),  
no signature of ejecta-companion interaction in some SNe Ia (e.g., Olling et al. 2015),
no detection of early  radio emission (e.g., Hancock et al. 2011; Horesh et al. 2012), 
and no absolute evidence of surviving companion stars of SNe Ia 
(e.g., Badenes et al. 2007; Kerzendorf et al. 2009, 2012, 2013, 2017b; 
Schaefer \& Pagnotta 2012; Edwards et al. 2012; Graham et al. 2015), etc.

In addition,  many observational evidence indicates that 
SN 2011fe may result from the merging of two WDs, which is one of the  nearest normal SNe Ia 
discovered by the Palomar Transient Factory soon after its explosion ($<1$\,d) and quickly followed by many wavebands
(e.g.,  Li et al. 2011b; Nugent et al. 2011; Brown et al. 2012; Bloom et al. 2012; Horesh et al. 2012; 
Liu et al. 2012b; Shappee et al. 2013, 2017; Chomiuk  2013; Parrent et al. 2014; Lundqvist et al. 2015).
Furthermore, many double WDs have been suggested to be
progenitor candidates of SNe Ia  (for more discussions see Sect. 4.3). 
Additionally,  Shen et al. (2018)  recently argued that prompt detonations of sub-${M}_{\rm Ch}$ WD in double WDs
can account for the observations of sub-luminous SNe Ia in old stellar populations.

However, the DD model has difficulties to explain the similarities of most SNe Ia 
as the WD explosion mass has a relatively wide range.
Meanwhile, a fundamental challenge for this model  is that
the merger of double WDs may result in the formation of  neutron stars through accretion-induced collapse
but not thermonuclear explosions (e.g., Saio \& Nomoto 1985, 1998; Nomoto \& Iben 1985; Kawai et al. 1987; 
Timmes et al. 1994;  Shen et al. 2012; Schwab et al. 2016).  
Due to a high $\dot{M}_{\rm acc}$ during the merger or in the post-merger cooling
stage (e.g., Yoon et al. 2007),  off-center carbon burning may happen on the surface of the CO WD, 
which likely converts CO WDs to ONe WDs through an inwardly propagating carbon flame but not SNe Ia. 
Note that Yoon et al.  (2007)
argued that the accretion-induced collapse may be avoided for a certain
range of parameters when  the rotation of the WDs is considered (see also Piersanti et al. 2003).

\subsection{The violent merger scenario}

It has been  suggested that the accretion-induced collapse 
may be avoided when the coalescence process of double WDs is violent, known as  the violent merger scenario;
a prompt detonation is
triggered when the merging continues, leading to an SN Ia
explosion (see Pakmor et al. 2010, 2011, 2012).
Pakmor et al. (2010) found that the violent merger of double WDs with almost equal-masses  ($\sim$$0.9\,M_{\odot}$)
is compatible with the low peak luminosity of SN 1991bg-like objects;
although the predicted light curves are too broad owing to the large ejecta mass,
the low expansion velocities and synthesized spectra match well with the observed SN 1991bg-like objects.
Following a 3D simulation for the violent merger of double WDs 
with masses of 1.1 and $0.9\,M_{\odot}$, 
Pakmor et al. (2012) suggested that 
the violent merger scenario may also explain the properties of normal SNe Ia.

The mass ratio of  double WDs has a great influence on the outcomes of the WD mergers.
Pakmor et al. (2011) argued that the minimum critical mass
ratio for double WD mergers to form SNe Ia is $\sim$0.8. 
The absolute SN Ia brightness in this scenario are mainly determined by the mass of the primary WD 
as the less massive WD will be totally destroyed during the merging (e.g., Ruiter et al. 2013).
It is still under debate whether the violent merger scenario  can really produce SNe Ia or not
(see, e.g., van Kerkwijk et al. 2010;
Taubenberger et al. 2013; Kromer et al. 2013; Moll et al. 2014;
Raskin et al. 2014; Fesen et al. 2015; Tanikawa et al. 2015;
Chakraborti et al. 2015; Sato et al. 2016; Bulla et al. 2016).

\subsection{Evolutionary scenarios  and parameter space}

\begin{figure}
\begin{center}
\epsfig{file=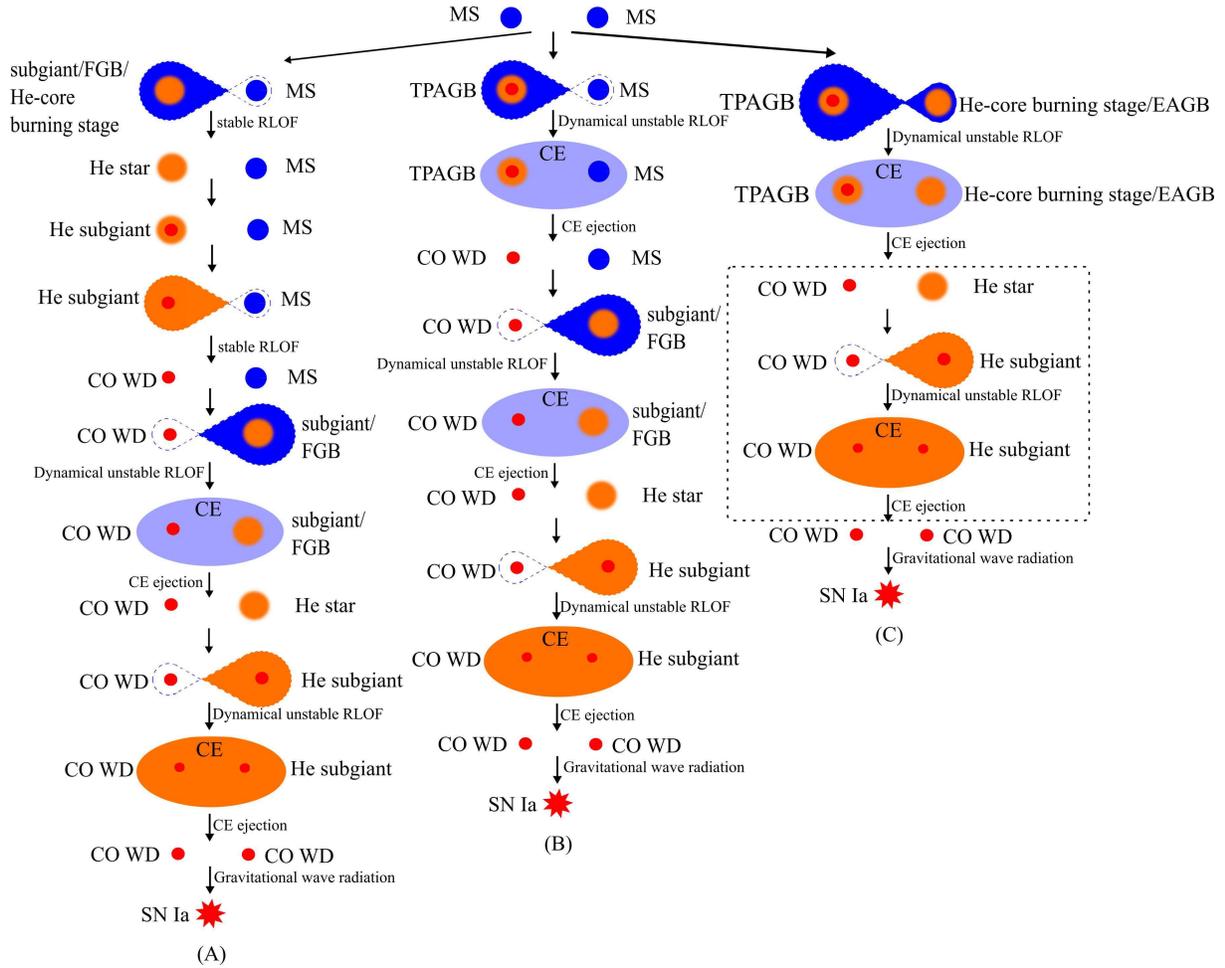,width=16cm} \caption{The common-envelope (CE) ejection scenarios to double WDs that  can form SNe Ia.
In Scenario C, some cases will not experience the second CE ejection (dashed box).}
\end{center}
\end{figure}

For the DD model,  it has been suggested that  there are three binary evolutionary
paths to form double WDs and then produce SNe~Ia; these formation paths can be named as  
the common-envelope (CE) ejection scenario
as all double WDs originate from the CE ejection process before the formation of DD systems
(see Fig. 11; e.g., Han 1998; Postnov \& Yungelson 2006; Toonen et al. 2012; 
Yungelson \& Kuranov 2017; Liu et al. 2018b).    
For Scenario A, the initial parameters of the primordial binaries are in the range of
$M_{\rm 1,i}\sim4.5$$-$$9.0\,M_\odot$, $q=M_{\rm 2,i}/M_{\rm 1,i}\sim0.2-0.8$, and $\log P^{\rm i}\rm(d) \sim 0.5-3.0$.
For Scenario B, the initial binary parameters are in the range of
$M_{\rm 1,i}\sim3.0$$-$$6.5\,M_\odot$, $q=M_{\rm 2,i}/M_{\rm 1,i}\sim0.3-0.9$, and $\log P^{\rm i}\rm(d) > 3.0$.
For Scenario C, the initial binary  parameters  are in the range of
$M_{\rm 1,i}\sim3.0$$-$$6.5\,M_\odot$, $q=M_{\rm 2,i}/M_{\rm 1,i}>0.9$, and $\log P^{\rm i}\rm(d)>3.0$.
In Scenario C, the double WDs can be also formed after the first CE ejection directly in some cases.
Among the three scenarios, SNe Ia are mainly produced by Scenarios A and B, in which each scenario 
contributes to about 45\% SNe Ia  (see Liu et al. 2018b).

Aside from the CE ejection scenario above,  Ruiter et al. (2013)  recently suggested
an important stage to the modeling of double WDs in the context of
violent mergers, namely a stage where the first-formed CO WD
increases its mass by accretion of helium from a He subgaint star, which is known as the WD+He subgiant scenario;
in this scenario the mass-transfer before the formation of double WD systems is dynamically stable, which can be also named as 
the stable mass-transfer scenario compared with the CE ejection scenario (see also Liu et al. 2016, 2018b). 
The  WD+He subgiant scenario allows the formation of significantly more massive primary CO WDs and thus
more massive double WDs, which can greatly enhance the SN Ia rate through the DD model if double WD mergers  can actually produce SNe Ia.
After considering the WD+He subgiant scenario, Liu et al. (2018b) found that the delay time distributions form the DD model 
is comparable with the observed results, and  that the violent mergers through the DD model may contribute to up to 16\% of all SNe Ia.

\begin{figure}
\begin{center}
\epsfig{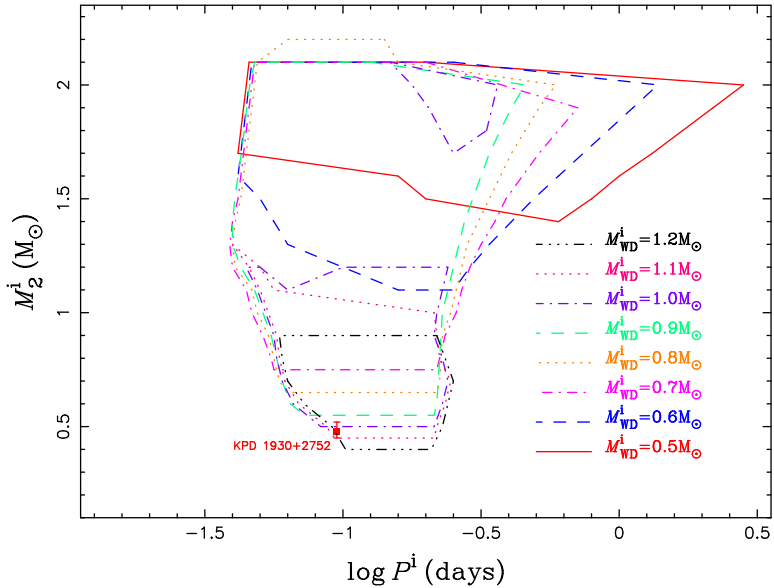} \caption{Initial parameter space of WD+He star systems for producing SNe Ia based on the DD model.  
The square with error bars shows the location of a WD+sdB star system KPD 1930+2752. 
The data points of these contours  are  from  Liu et al. (2018b).}
\end{center}
\end{figure}

The WD+He subgiant scenario has a significant contribution to the formation of massive double WDs. 
Fig. 12 shows the initial parameter space of WD+He star systems for producing SNe Ia based on the 
DD model in the $\log P^{\rm i}-M^{\rm i}_2$ plane with different ${M}^{\rm i}_{\rm WD}$. 
The contours turn to move to upstairs for lower ${M}^{\rm i}_{\rm WD}$, resulting from the 
assumption  that the total mass of double WDs needs to be $\geq$${M}_{\rm Ch}$ for producing 
SNe Ia. The WD+He star systems  outside the contours cannot produce SNe Ia through the DD model 
(for more details see Sect. 2.3 of Liu et al. 2018b). The parameters of KPD 1930+2752 are 
located in the parameter space of WD+He star systems for producing SNe Ia through the DD model 
(see Fig. 12), indicating that it is a progenitor candidate of SNe Ia.

\begin{figure}
\begin{center}
\epsfig{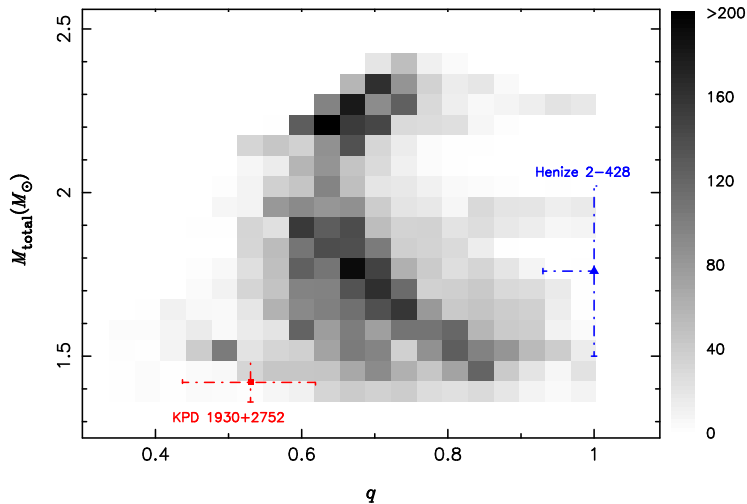} \caption{Density distribution of the masses of double WDs that can 
form SNe Ia in the mass-ratio and total mass plane.  
The blue filled triangle with error bars represent the DD core of a planetary nebula Henize 2-428, and the red filled square with error bars
shows a double WDs that originates from KPD 1930+2752. 
The data points of the  density distribution are from  Liu et al. (2018b).}
\end{center}
\end{figure}

The outcomes of WD mergers are determined by their mass-ratio and total mass (e.g., Pakmor et al. 2010; Sato et al. 2016). 
Fig. 13 shows  the density distribution of the masses of WD mergers that can form SNe Ia in the mass-ratio and total mass plane.
The total masses of double WDs for producing SNe Ia have a wide distribution ranging from 1.378$-$2.4\,$M_\odot$, 
and the mass-ratio are mainly located in the range of 0.6$-$0.8. 
The density distribution can be divided into two parts:  (1) The mass ratio decreases with the total mass in the less-massive part that originates 
from the WD+He subgiant scenario or the CE ejection scenario.  (2) The mass ratio increases with the total mass in the massive part that
mainly originates from the WD+He subgiant scenario.  The parameters of KPD 1930+2752 and Henize 2-428  are located in the mass-ratio and total mass plane.
KPD 1930+2752 originates from the WD+He subgiant scenario, whereas Henize 2-428 originates from the CE ejection scenario  (for more discussions see Sect. 4.3).
 
\subsection{Progenitor candidates}

KPD 1930+2752 and Henize 2-428 are two progenitor candidates of SNe Ia through the DD model. 
(1) KPD 1930+2752 is a WD+sdB system with  an   orbital period of $\sim$2.28\,h (see Maxted et al. 2000).
The mass of the sdB star is  $\sim$0.45$-$0.52\,$M_\odot$ and the total mass of the system  is  $\sim$1.36$-$1.48\,$M_\odot$ (see Geier et al. 2007).
Liu et al. (2018b) recently suggested that KPD 1930+2752 will not experience mass-transfer  until the formation of a double WDs; 
it takes $\sim$200\,Myr for KPD 1930+2752 to form a double WDs. 
After the formation of double WDs,  KPD 1930+2752 will merge in $\sim$4\,Myr.
(2) Henize 2-428 is a planetary nebula with a
DD core that has a total mass $\sim$1.76\,$M_\odot$ and mass-ratio $\sim$1 
with an orbital period of $\sim$4.2\,h (see Santander-Garc\'ia et al. 2015), 
which is  a strong progenitor candidate of SNe Ia through the violent merger scenario.

Recently, some other
double WDs have been found, which may have the total mass close to ${M}_{\rm Ch}$ and
likely merge in the Hubble-time.  For example, WD 2020-425  (e.g., Napiwotzki et al. 2007),
V458 Vulpeculae  (e.g., Rodr\'{\i}guez-Gil et al. 2010),
SBS 1150+599A  (e.g., Tovmassian et al. 2010), and GD687  (e.g., Geier et al. 2010), etc. 
Kawka et al. (2017) recently argued that NLTT 12758 is a super-${M}_{\rm Ch}$ double WD system, but its merging timescale
is larger than the Hubble time.
So far, there are some systematic surveys for searching double WDs,
for example, ESO SN~Ia Progenitor Survey (SPY; e.g., Koester et al. 2001; Geier et al. 2007;
Napiwotzki et al. 2004; Nelemans et al. 2005) and the SWARMS survey  (see Badenes et al. 2009b).
In addition, a substantial population of double WDs may be obtained by 
Gaia (e.g., Carrasco et al. 2014; Toonen et al. 2017).
Before Gaia DR2 is released, the GPS1 proper motion catalogue could be one of most potential catalogs 
to obtain a substantial population of double WDs owing to its accurate kinematic and photometric informations (see Tian et al. 2017).
Furthermore, this type of WD binaries is an important   kind  of gravitational wave sources in our Galaxy 
(e.g., Yu \& Jeffery 2010, 2015; Liu et al. 2012c; Liu \& Zhang 2014).
Kremer et al. (2017) recently predicted that about 2700 double WD gravitational wave sources will be observable by LISA in our Galaxy.

\section{The sub-Chandrasekhar  mass model}

In this model, a CO WD accumulates a substantial He-shell by mass accretion 
with a total mass below ${M}_{\rm Ch}$, 
the explosion of which is triggered by the detonation at the bottom of He-shell;
one detonation propagates outwardly via the He-shell, whereas another inwardly propagating pressure wave compresses 
the CO-core and leads to  carbon ignition,  which is known as the double-detonation model 
(e.g., Nomoto 1982a; Woosley et al. 1986; Livne 1990; Branch et al. 1995; H\"{o}flich \& Khokhlov 1996). 
The minimum WD mass for this model might be $\sim$0.8\,$M_\odot$ as the detonation of 
the WD may be not triggered for lower mass (e.g., Sim et al. 2012).

It has been suggested that the sub-${M}_{\rm Ch}$ model might explain
sub-luminous SN 1991bg-like objects   (e.g., Branch et al. 1995; Dhawan et al. 2017; 
Blondin et al. 2018),\footnote{According to the sub-${M}_{\rm Ch}$ double-detonation model, 
Liu et al. (2017) suggested that the merging of a CO WD with a He-rich WD (a He WD or a hybrid HeCO WD) 
can roughly reproduce the rates of SN 1991bg-like objects.}
and that this model 
may account for at least some substantial fraction of the observed SN Ia rates if this model can really form SNe Ia
(e.g., Ruiter et al. 2009, 2011).  
Fink et al. (2010) argued that the double-detonation explosion in sub-${M}_{\rm Ch}$ WDs 
could be robust, even resulting in the formation of normal SNe Ia.
According to multiwavelength radiation transport simulations, 
Goldstein \& Kasen (2018) recently suggested that the sub-${M}_{\rm Ch}$ model 
can reproduce the entirety of the width-luminosity relation of the observed SNe Ia.
However,  this model still fails to explain
many of the main properties of the observed SNe Ia so far, and 
it is still  \textbf{uncertain} that this model can really interpret which known SNe Ia   (e.g., 
Nugent et al. 1997; Bildsten et al. 2007; Fink et al. 2007, 2010; Shen \& Bildsten 2009; Sim et al. 2010; 
Kromer et al. 2010; Ruiter et al. 2011; Woosley \& Kasen 2011).

Jiang et al. (2017) recently observed a hybrid SN Ia (SN 2016jhr), 
which has a light curve like normal SNe Ia but with strong titanium absorptions like sub-luminous events;
this SN Ia has a prominent but red optical flash at $\sim$0.5\,d after the SN explosion. 
Jiang et al. (2017) suggested that the early flash of such a hybrid SN Ia may be naturally 
interpreted by a SN explosion triggered by the detonation of a thin He-shell. 
Sarbadhicary et al. (2017) recently studied two
young SN remnants (SN 1885A and G1.9+0.3), which are the most recent SN Ia remnants in the Local Group.
They argued that  SN 1885A is consistent with 
the sub-${M}_{\rm Ch}$ explosion model, 
and both ${M}_{\rm Ch}$ and sub-${M}_{\rm Ch}$ explosion  models   are likely for  the SN remnant
G1.9+0.3.

\subsection{Parameter space}

\begin{figure}
\begin{center}
\epsfig{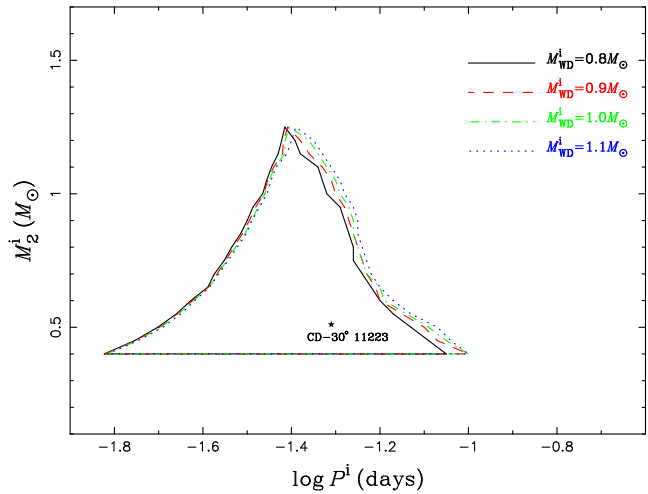} \caption{Initial parameter space of WD+He star systems for producing SNe Ia 
based on the sub-${M}_{\rm Ch}$ model. The filled star shows the location of a WD+He star system CD$-$30$^{\circ}$\,11223.  
The data points of these contours are from Wang et al. (2013b).}
\end{center}
\end{figure}

In the  sub-${M}_{\rm Ch}$ model, a  CO WD with mass below ${M}_{\rm Ch}$ can  accrete material from a non-degenerate He star.
Fig. 14 shows the initial parameter space of CO WD+He star systems for producing SNe Ia 
based on the sub-${M}_{\rm Ch}$ model. 
In this figure, 
systems beyond the right boundary of the space will experience a high mass-transfer rate
as the He star evolves to the He-shell burning phase that is not suitable to trigger double-detonation, and
the left boundaries  are determined by the
condition that 
RLOF happens when the He donor  is on the zero-age MS phase.
The lower boundaries are set by the condition that  the mass-transfer rate
is high enough to form  a critical He-shell to trigger a detonation on the surface of the WD.

Foley et al.\ (2013) proposed  a distinct sub-class
of sub-luminous SNe Ia, named  type Iax
SNe  that  include SNe resembling the prototype object SN 2002cx 
(e.g., Li et al.\ 2003; Jha 2017; Barna et al. 2017; Lyman et al. 2018; Singh et al. 2018).
Wang et al. (2013b) estimated that the  
Galactic SN Ia rate  from the  sub-${M}_{\rm Ch}$ model  is 
in good agreement with the measured rates of  type Iax SNe, and that this model can reproduce 
the delay time distributions and the luminosity distribution of  type Iax SNe. 
The binary parameters of  CD$-$30$^{\circ}$\,11223 (a WD+He star system) are located in the parameter space of 
WD+He star systems for producing SNe Ia (see Fig. 14), which means that this binary is a
progenitor candidate of SNe Ia (for more discussions see Sect. 5.2).
For the sub-${M}_{\rm Ch}$ model, the ignition mass for the accumulated He-shell is still not well determined, 
which may depend on $\dot{M}_{\rm acc}$ and
change with the temperature of the WD and the CO-core mass 
(e.g., Iben \& Tutukov 1989; Bildsten et al. 2007; Shen \& Bildsten 2009; 
Ruiter et al. 2014; Neunteufel et al. 2016).

\subsection{Progenitor candidates}

CD$-$30$^{\circ}$\,11223  has been identified
as a CO WD+He star system with a $\sim$1.2\,h orbital period, including a
$\sim$$0.76\,M_{\rm\odot}$ WD and a
$\sim$$0.51\,M_{\rm\odot}$ He star
(see, e.g.,Vennes et al.\ 2012; Geier et al.\ 2013). 
Angular momentum loss from the  short orbital system is large  due to gravitational wave radiation.
After $\sim$36\,Myr, the He star will start to fill its
Roche lobe when it is still in the He MS stage.
CD$-$30$^{\circ}$\,11223 may produce an SN Ia through the sub-${M}_{\rm Ch}$ model in the subsequent
evolution (e.g., Wang et al. 2013b; Geier et al.\ 2013).
The mass donor star in the sub-${M}_{\rm Ch}$ model would survive after the SN explosion.
Geier et al. (2013) suggested that CD$-$30$^{\circ}$\,11223 and the hypervelocity He star US 708 might show
 two different evolutionary phases (i.e., progenitor and remnant) linked by an SN Ia explosion.

Motivated by the discovery of CD$-$30$^{\circ}$\,11223, Kupfer et al. (2017) started a search for ultracompact 
post-CE binaries based on the Palomar Transient Factory  survey.
Kupfer et al. (2018) recently reported the discovery of an ultracompact WD+sdOB system OW J0741 with an orbital period of 44\,min
based on the OmegaWhite survey, including a $0.72\pm0.17\,M_{\rm\odot}$  WD and a $0.23\pm0.12\,M_{\rm\odot}$  sdOB star. 
They argued that this binary will either end up as a stably mass-accreting AM CVn system or merge to form an R CrB star eventually.
So far, OW J0741 is  the most compact WD+hot subdwarf system known. 
It is noting that Luo et al. (2016) is searching for hot subdwarf binary systems using LAMOST data.

\section{Other possible models}

Aside from the SD model, the DD model and the sub-${M}_{\rm Ch}$ model above, 
some potential progenitor models have been proposed to explain the observed diversity among SNe~Ia.
For example, the hybrid CONe WD model,
the core-degenerate model, 
the double WD  collision model, 
the spin-up/spin-down model, 
and the model of WDs near black holes,  etc 
(for recent reviews see Wang \& Han 2012;  Maoz et al. 2014; Soker 2018).

\subsection{The hybrid CONe WD model}

Hybrid CONe WDs have been suggested to be possible progenitors of SNe Ia, which
has an unburnt CO-core surrounded by a thick ONe-shell (e.g., Denissenkov et al. 2013, 2015; Chen et al. 2014b).
These hybrid WDs can easily grow in  mass to ${M}_{\rm Ch}$ by accreting matter,
which could increase the  rates  of SNe Ia if they can really form SNe Ia. 
Denissenkov et al. (2015) recently argued that hybrid WDs could reach a state of explosive
carbon ignition though it depends on some mixing assumptions and the convective Urca process.
It has been suggested that carbon abundance in hybrid CONe WDs is lower than that of CO WDs (e.g., Denissenkov et al. 2015). 
Therefore,  hybrid CONe WDs are expected to form SNe Ia with lower peak luminosity and explosion energy, and thus  
a relatively low ejecta velocity.  

A hybrid CONe WD could increase its mass to ${M}_{\rm Ch}$
by accreting H-rich material from a MS star (the CONe WD+MS scenario; see Meng \& Podsiadlowski 2014) 
or from a He star (the CONe WD+He star scenario; see Wang et al. 2014b).
Wang et al. (2014b) found that SNe Ia from the CONe WD+He star scenario 
could be as young as $\sim$28\,Myr, which are the youngest SNe Ia ever predicted. 
Wang et al. (2014b) also suggested that the CONe WD+He star scenario could provide an alternative way for producing type Iax SNe  like SN 2012Z
that may be an explosion of a WD accreting material from a He star.
By calculating the hydrodynamical stage of the explosion of  CONe WDs with ${M}_{\rm Ch}$,  Bravo et al. (2016) recently claimed that
CONe WDs cannot explain the properties of normal SNe Ia though they may form SNe Iax.
For more discussions on this model, see, e.g.,  Willcox et al. (2016) and Brooks et al. (2017b)

\subsection{The core-degenerate model}
In this model, SNe Ia are produced 
at the final stage of common-envelope (CE) evolution through the merging
of a CO WD with the hot CO-core of an AGB  star
 (e.g.,  Kashi \& Soker 2011; Ilkov \& Soker 2012; Soker et al. 2013, 2014).
It has been suggested that this model provides an alternative way to form super-luminous 
SNe Ia (e.g.,  Kashi \& Soker 2011; Ilkov \& Soker 2012) and
some SNe Ia with circumstellar material like PTF 11kx (e.g., Soker et al. 2013). 
This model was also used to explain the formation of SNe 2014J and 2011fe  (e.g., Soker et al. 2014; Soker 2015).
According to 3D smoothed particle hydrodynamics simulations,
Aznar-Sigu\'{a}n et al. (2015) recently argued that a massive CO
WD can be  produced through this merging process,  resulting in an SN Ia explosion finally. 
In order to search for the surviving companion star of Kepler's SN, 
Ruiz-Lapuente et al. (2018) recently surveyed the remnant of this SN and
suggested that Kepler's SN could originate either from the core-degenerate model 
or from the DD model based on the strong limits placed on luminosity,   

However, the rates of SNe Ia from the core-degenerate model are still not well determined.
Ilkov \& Soker (2013) argued that this model  can reproduce
the observed rates of all SNe Ia based on a simplified BPS  code.
Due to more careful treatment of mass-transfer process, 
Wang et al. (2017b) suggested that the Galactic rates of SNe Ia from this model are no more than 20\% of all SNe Ia,
mainly contributing to the observed  ones with short and intermediate delay times.
Wang et al. (2017b)  estimated that SNe Ia with circumstellar material from the core-degenerate model can 
account for 0.7$-$10\% of all SNe Ia,  which can explain the observed number of SNe Ia  like PTF 11kx. 
At present, it seems that the  core-degenerate model cannot be excluded as a viable way for the production of SNe Ia.
Soker (2018) recently summarized the properties of different progenitor models and
made detailed comparisons between the core-degenerate model and other  models  (see also Tsebrenko \& Soker 2015).
 
\subsection{The double WD collision model}

This model is a variant of the DD model, which involves the direct collisions of two WDs 
(e.g., Raskin et al. 2009, 2010; Katz \& Dong 2012; Kushnir et al. 2013). 
It had been commonly assumed that double WD collisions only occurred in dense stellar environments 
such as globular clusters and they would thus be a negligible fraction of SNe Ia (e.g., Raskin et al. 2009). 
In the last few years, it has been realized that the rate of double WD mergers or collisions can be significantly 
enhanced due to few-body dynamics in field multiples (see, e.g., Thompson et al. 2011; Katz \& Dong 2012; Pejcha et al. 2013). 
The study by Katz \& Dong (2012) showed that the non-secular effects of Lidov-Kozai mechanism 
can enhance the double WD collision rates by several orders of magnitude than previously thought 
and raised the possibility that the collision rate might be on the same order of magnitude with SN Ia rate. 
Kushnir et al. (2013) demonstrated successful detonations of double WD collisions, 
which could produce $^{56}$Ni masses spanning the whole range of observed SNe Ia luminosity function from SN 1991T-type events to SN 1999bg-like events. 
Dong et al. (2015) discovered double-peaked  [CoIII] lines in the nebular phase spectra of 3 out of 20 SNe Ia and suggested 
that SNe Ia with intrinsic bi-modality in $^{56}$Ni may be common among sub-luminous ones ($\sim$40\% of all SNe Ia), 
which is naturally expected from direct collisions of two WDs due to two centers of detonations. 

One major open question for the double WD collision model is whether post-MS stellar evolution can produce 
the sufficient amount of double WDs in suitable multiple stellar systems (see Katz \& Dong 2012; Toonen et al. 2018). 
Further works on the effects of stellar evolution and stellar multiplicity 
(e.g., Klein \& Katz 2017) may help to advance our understanding of the rate issue of the collision model.

\subsection{The spin-up/spin-down model}
In this model, a WD can be spun up by mass accretion from its donor, which can increase its mass above ${M}_{\rm Ch}$;
the WD likely needs a spin-down time before it explodes as an SN Ia (e.g., Di Stefano et al. 2011; Justham 2011).
This model is a variant of the SD model, which provides a way to reproduce the observed
similarities and diversity  among SNe~Ia.
Due to the spin-up of the WD, the SD model can naturally explain 
the observed super-luminous SNe~Ia (e.g., Hachisu et al. 2012; Wang et al. 2014a; Benvenuto et al. 2015).
By  considering  the effect of rotation on accreting WDs,
Wang et al. (2014a) predicted that 2\% of SNe Ia from the SD model happen with WD explosion masses 
$\geq$$2\,M_\odot$, which is broadly comparable
with these super-luminous ones; these super-luminous  events require the initial WD mass to be $>$$1.0\,\rm M_{\odot}$.

Importantly, after considering the spin-down time, 
the SD model could be  consistent with the observed properties of most SNe Ia, in particular for the absence of H line in the late-time spectra.
However, the spin-down time is still quite uncertain. 
Meng \& Podsiadlowski (2013) recently argued that the upper limit of the spin-down time is a few $10^{7}$\,yr for progenitor systems that include a RG donor.
For more discussions about this model, see, e.g.,  Yoon \& Langer (2004), Chen \& Li (2009) and Ghosh \& Wheeler (2017).
 
\subsection{The  model of WDs near black holes}
 In this model, SNe Ia are produced by relativistic enhancements of the WD self-gravity 
 when the WD passes near a black hole; this
relativistic compression can make the central density of the WD
exceed the threshold for pycnonuclear reactions,  leading to 
thermonuclear  explosions (see Wilson \& Mathews 2004). 
The observed ``mixed-morphology'' of the Sgr A
East SN remnant in the Galactic center might be explained by this mechanism (see Dearborn et al. 2005).
Rosswog et al. (2009) speculated that 
such encounters may be frequent   in the center of dwarf galaxies or
globular clusters that host an intermediate-mass black holes  (see also Rosswog et al. 2008).
Note that the WD+black hole tidal disruption may lead to different events, depending on 
the mass of the black hole, mass of the WD, and pericenter orbital radius (see Kawana et al. 2017).
According to high-resolution simulations,  Tanikawa (2018) recently suggested that  WDs 
near black holes can explode as SNe Ia  through the tidal double-detonation mechanism (see also Tanikawa 2017).
I estimate that  the SN~Ia
rate from this model may be relatively low as the  low possibility of a
WD passing close to a black hole, and this model might only explain SNe Ia nearby black holes.

\section{Summary}

Mass-accreting CO WDs are expected to form SNe Ia when they grow in mass close to ${M}_{\rm Ch}$.
Recent studies on mass-accreting CO WDs (including H- and He-accreting WDs) are reviewed, which is
important for understanding the mass increase of the WD.
Currently, the most studied SN Ia progenitor models are the SD model, the DD model
and the sub-${M}_{\rm Ch}$ model. I review recent progress on these progenitor models, 
including the initial parameter space for producing SNe Ia, 
the binary evolutionary paths to SNe Ia, the progenitor candidates of SNe Ia, 
the possible surviving companion stars of SNe Ia,
and some observational constraints, etc. 
The issue of the progenitors of SNe Ia is still poorly understood. 
It is still no single progenitor model that can reproduce all the observational features and full diversity of SNe Ia.
So far, it seems that two or more progenitor models,  including some other potential progenitor models, 
may contribute to the  observed  diversity among SNe Ia,  although the fraction of SNe Ia from each model is really uncertain.
To provide further constraints on the issue of SN Ia progenitors, 
large samples of well-observed SNe~Ia and progenitor candidates are needed, 
and new progress on the theoretical side is  expected. 
Additionally, a large number of ongoing surveys from ground and space are  searching  for more SNe Ia,\footnote{For example,
the Panoramic Survey Telescope and Rapid Response System (Pan-STARRS),
the Gaia Astrometric Mission,
the Asteroid Terrestrial-impact Last Alert System (ATLAS), 
the Catalina Real-Time Transient Survey  (CRTS), 
the All-Sky Automated Survey for Supernovae (ASAS-SN), 
the OGLE-IV wide field survey,
the PMO-Tsinghua Supernova Survey (PTSS), 
the SkyMapper,
the Dark Energy Survey, 
the Palomar Transient Factory (PTF),
and the THU-NAOC Transient Survey (TNTS), etc
(e.g., Altavilla et al. 2012; Parrent et al. 2014; Zhang et al. 2015; Chambers et al. 2016; for more, see http://www.rochesterastronomy.org).}
which may build the connections between SN Ia progenitors and the observed properties of SN explosions.

\begin{acknowledgements}
BW acknowledges the referee for the valuable comments that helped me
to improve this review. 
BW also thanks Zhanwen Han, 
Thomas Marsh, Noam Soker, Or Graur, Tyrone Woods,
Subo Dong,  Xiangcun Meng, Xuefei Chen, Zhengwei Liu,  Jujia Zhang, 
Hailiang Chen, Ataru Tanikawa and  Dongdong Liu
for their helpful comments and suggestions.
This work is supported by the National Basic Research Program of China (973 programme, 2014CB845700), 
the National Natural Science Foundation of China (Nos 11673059, 11521303 and 11390374),  
the Chinese Academy of Sciences (Nos KJZD-EW-M06-01 and QYZDB-SSW-SYS001),
and the Natural Science Foundation of Yunnan Province (Nos 2013HB097 and 2017HC018).

\end{acknowledgements}

\label{lastpage}
\end{document}